# DaYu-TC: a global–regional nested AI system for tropical cyclone forecasting

Runzepu Hou[1,2], Feng Zhang[1,2*], Xin Wu[1,2], Xujun Wei[1,2], Fenghua Ling[3], Kan Dai[4], Gaozhen Nie[4,5], Leiming Ma[6], Renhe Zhang[1]

[1] Key Laboratory of Polar Atmosphere-Ocean-Ice System for Weather and Climate of Ministry of Education / Shanghai Key Laboratory of Ocean-Land-Atmosphere Boundary Dynamics and Climate Change, Department of Atmospheric and Oceanic Sciences & Institutes of Atmospheric Sciences, Fudan University, Shanghai 200438, China

[2] Key Laboratory for Information Science of Electromagnetic Waves, Ministry of Education, College of Future Information Technology, Fudan University, Shanghai 200438, China

[3] Shanghai Artificial Intelligence Laboratory, Shanghai 200232, China

[4] National Meteorological Center, Beijing 100081, China

[5] Xiong'an Institute of Meteorological Artificial Intelligence, Hebei 070001, China

[6] Shanghai Typhoon Institute, China Meteorological Administration, Shanghai 200030, China

* Correspondence: Feng Zhang; Email: fengzhang@fudan.edu.cn; Telephone: +86 13310066269

## Abstract

Accurate tropical cyclone intensity forecasts are essential for disaster preparedness, yet persistent intensity underestimation limits the value of artificial intelligence (AI) weather models for predicting these destructive storms. Here we present DaYu-TC, a global–regional nested system that links the preservation of strong-wind and low-pressure signals with high-resolution regional prediction. Its global and regional models share an architecture and extreme-aware training objectives. The regional model combines evolving local states with global forecasts valid at each target time, connecting large-scale environmental information with finer-scale cyclone evolution. In an independent evaluation of 16 western North Pacific tropical cyclones in 2023, the global model retains competitive conventional forecast skill while reducing five-day maximum-wind root-mean-square error (RMSE) by 6–22% relative to other evaluated global AI models. Relative to the global model, regional nesting reduces maximum-wind RMSE by 22.0–33.1% and minimum-pressure RMSE by 28.6–39.8% across 12–120 h.

The complete system outperforms concurrent operational high-resolution forecasts from the European Centre for Medium-Range Weather Forecasts for minimum pressure at 12–120 h and maximum wind speed at 48–96 h, with respective RMSE reductions of 26% and 12% at 72 h. Without retraining, the same regional model enhances three other global AI models, reducing overall maximum-wind RMSE by 22.6–28.6%. Five-day forecasts take under one minute on a single GPU. These results establish a transferable multiscale approach to reducing cyclone intensity underestimation, extending advances in AI weather forecasting towards more accurate prediction of high-impact weather events.

## 1. Introduction

Tropical cyclones (TCs) pose severe threats to human life, infrastructure and livelihoods [1,2]. Accurate forecasts of their tracks and intensities are essential for timely warnings and effective disaster preparedness [3,4]. In recent years, global artificial intelligence (AI) weather models, including Pangu-Weather and GraphCast, have demonstrated high forecast skill and computational efficiency in medium-range forecasting [5–8]. However, improvements in large-scale forecast skill have not fully translated into more accurate TC intensity forecasts [9]. Evaluations show that many global AI models predict TC tracks skilfully but systematically underestimate intensity [10,11]. Their intensity forecasts are less accurate than those of operational numerical models and often show early weakening and a persistent negative bias [12]. Improving the representation of TC intensity and near-centre structures while retaining large-scale forecast skill is therefore a key challenge for AI prediction of high-impact weather events [13].

At the global-model level, reducing the attenuation of extreme anomalies offers a route to improved TC intensity forecasts. Training objectives dominated by average gridpoint errors can smooth local extremes and spatial gradients, with further attenuation at longer lead times [14,15]. Lopez-Gomez et al. improved global high-temperature forecasts using an extreme-focused loss function and reduced forecast smoothing at longer lead times [16]. Subich et al. enhanced predictions of extreme surface winds and TC intensity by fine-tuning GraphCast with a modified spherical harmonic loss [14]. Complementary advances include diffusion-based refinement of extreme winds and precipitation [17], frequency-domain constraints for sharper precipitation fields [18] and contrastive learning for extreme-temperature forecasting [19]. These studies demonstrate the value of learning objectives and representations tailored to weather extremes. Building on these advances, a central priority for TC-focused global modelling is to jointly preserve strong-wind and low-pressure anomalies throughout multi-step forecasts and translate this preservation into sustained intensity gains.

In addition to attenuation during prediction, biases in training data affect the representation of TC intensity and inner-core structure [20]. ERA5, widely used to train global AI weather models, typically underestimates TC intensity because its limited resolution constrains near-centre wind maxima and fine-scale structures [21]. Global–regional modelling provides a means of incorporating finer-scale information, building on the use of global forecasts to constrain regional numerical prediction [22,23]. Recent AI approaches connect global and regional prediction in different ways. StormCast uses large-scale forcing to drive kilometre-scale regional forecasts [24], while OneForecast couples 1.5° global forecasts with 0.25° regional predictions [25]. Stretched-grid models retain global coverage while increasing resolution over selected regions [26,27]. For TC prediction, FuXi-TC uses high-resolution Weather Research and Forecasting (WRF) simulations as training targets to post-process global AI forecasts, refining wind and precipitation fields [28]. HITS combines regional states from a dedicated

TC reanalysis dataset with future forecasts from the AI Forecasting System (AIFS) [29]. These studies establish regional prediction as a promising approach to strengthening TC intensity and structural representation under large-scale environmental constraints.

Taken together, previous studies have advanced global extreme-weather prediction and regional refinement, but their combined contribution to TC intensity forecasts requires systematic evaluation. It remains unclear whether preserving strong-wind and low-pressure anomalies in a global model can provide a stronger foundation for regional prediction, producing cumulative gains from global extreme representation to TC intensity. Moreover, existing regional approaches are commonly developed for a specific global driver or specialized high-resolution training dataset. The transferability of these gains to other global AI drivers without retraining warrants direct evaluation. Resolving these issues is essential for turning regional AI prediction from a model-specific component into a reusable enhancement for global forecasting systems.

Here we introduce DaYu-TC, a global–regional nested AI forecasting system designed to connect global extreme preservation, regional TC refinement and cross-driver transfer within a unified framework. It comprises the 0.25° DaYu-Global and 0.1° DaYu-Region models. DaYu-Global uses extreme-aware objectives and structure-enhancement modules to preserve strong-wind and low-pressure anomalies during multi-step forecasting. Initialized with the global model's weights, DaYu-Region combines consecutive high-resolution regional states with global forecasts at the target time to predict regional atmospheric fields at finer resolution. The same regional weights are further evaluated under multiple global AI drivers without retraining. DaYu-Global achieves the lowest maximum-wind errors among the evaluated global AI models from approximately 24 h onwards, while regional nesting further improves wind and pressure intensity forecasts and outperforms the European Centre for Medium-Range Weather Forecasts (ECMWF) Integrated Forecasting System high-resolution (IFS-HRES) forecasts over sustained lead-time ranges. The cross-driver gains demonstrate that regional refinement can operate as a transferable forecasting component rather than an enhancement tied to a specific global model.

## 2. Materials and methods

### 2.1. Forecasting framework

DaYu-TC combines global extreme preservation with high-resolution regional prediction through one-way nesting (Fig. 1). DaYu-Global predicts the next atmospheric state on a 0.25° grid from two states separated by 6 h. Its forecast valid at the next target time is cropped to the regional domain and interpolated to a 0.1° grid. DaYu-Region uses this background, two consecutive regional states and static geographic information to predict the next regional state. Both models advance autoregressively at 6 h intervals.

The models share a U-shaped Transformer with separate surface and upper-air encoders. The Trend Adapter represents temporal changes, Axial Attention captures long-range spatial relationships, and the Detail Adapter retains local high-frequency information. An Auxiliary Head is used to supply additional supervision for extreme structures during training. Architectural details are provided in the Supplementary Methods and Fig. S14.

### 2.2. Data and experimental design

DaYu-Global is trained on ERA5 reanalysis from 1990–2021, with 2022 used for validation and 2023 for testing [30]. DaYu-Region uses operational IFS analyses on a 0.1° grid over the western North Pacific (WNP; 75.0°–204.6°E, 10.0°S–54.8°N) [31]. Regional data from 2016–2022 are used for training, and 2023 is reserved for independent testing. Both models predict 69 channels: four surface variables and five upper-air variables at 13 pressure levels (Supplementary Table S2).

TC verification uses 16 WNP TCs in 2023 from NOAA's International Best Track Archive for Climate Stewardship (IBTrACS) [32,33]. Positions and times are taken from LAT, LON and ISO_TIME; intensity references are USA_WIND (US-agency 1-minute sustained wind estimates) and USA_PRES. The storm sample is shown in Supplementary Fig. S13.

### 2.3. Training strategy

Training first learns the atmospheric evolution and then emphasizes extreme structures. Both models use AdamW [34]. Basic training minimizes latitude-weighted mean absolute error. Extreme-focused training combines an extreme-aware field loss with spatial-gradient, extreme-value and auxiliary-task losses to emphasize strong winds and low-pressure structures. DaYu-Global undergoes 25 epochs of basic training, five epochs of extreme-focused training and five epochs of autoregressive fine-tuning using the replay-buffer strategy of FengWu [35].

DaYu-Region is initialized from the trained global weights and undergoes five epochs of regional adaptation followed by five epochs of extreme enhancement, with TC sample resampling in the second stage. Regional training uses a single-step 6 h objective, IFS analyses as targets and DaYu-Global forecasts as backgrounds. Detailed loss formulations and optimization settings are provided in the Supplementary Methods.

### 2.4. Forecast evaluation

Conventional forecast skill is assessed using latitude-weighted root-mean-square error (RMSE) and anomaly correlation coefficient (ACC) following WeatherBench 2 [36]. Global extreme events are identified where ERA5 anomaly magnitudes reach or exceed local calendar-day 95th-percentile thresholds, estimated from 1993–2016 using centred five-day windows [37]. Relative mean bias error (rMBE) measures the latitude-weighted anomaly-amplitude bias relative to the reference amplitude at these locations; negative values indicate attenuation.

TC forecasts are evaluated to 120 h against IBTrACS. Track error is the distance between forecast and best-track centres. Intensity errors are calculated separately for maximum 10 m wind speed and minimum mean sea-level pressure. Cross-driver experiments replace DaYu-Global backgrounds with Pangu-Weather, GraphCast and FengWu forecasts while retaining the same regional weights.

Inference benchmarks measure 20 successive 6 h steps on one NVIDIA A100 GPU, including global inference, background remapping and regional prediction for the complete system. Tracking procedures, metric definitions and benchmark details are provided in the Supplementary Methods and Table S1.

## 3. Results

We first evaluate global extreme-anomaly preservation and TC forecasts from DaYu-Global, then assess the additional intensity and structural gains from regional nesting, transfer across global drivers and computational cost. Conventional atmospheric-field evaluations are provided in the Supplementary Results.

### 3.1. Extreme preservation and TC forecasting by DaYu-Global

DaYu-Global shows consistently smaller biases in extreme wind and low-pressure anomaly amplitudes than the other evaluated models (Fig. 2). At 24, 72 and 120 h, its relative mean bias error (rMBE) is closer to zero for both zonal and meridional winds at 10 m and 850 hPa (U10, V10, U850 and V850), as well as mean sea-level pressure (MSLP), indicating better preservation of extreme-anomaly amplitudes across variables and lead times (Fig. 2b–d).

At 120 h, the rMBE values for DaYu-Global are −23.9%, −24.8%, −23.5% and −27.4% for U10, V10, U850 and V850, respectively, and −12.7% for MSLP (Fig. 2a). The absolute rMBE values for the four wind components are 7.5–8.9 percentage points lower than those of FengWu and 0.7–3.5 percentage points lower than those of FuXi, GraphCast and Pangu-Weather. For MSLP, the absolute rMBE is 1.6–4.0 percentage points lower than those of the other models. These results demonstrate that DaYu-Global better preserves the amplitudes of extreme wind and low-pressure anomalies during multi-step global forecasting.

These gains in extreme-anomaly preservation are accompanied by conventional forecast skill comparable to that of FuXi and FengWu across most evaluated variables on the 2023 test set (Supplementary Fig. S1). DaYu-Global also reproduces major global weather systems and their spatial patterns at extended lead times (Supplementary Fig. S2).

Building on the global extreme-anomaly evaluation, we assess TC track and intensity forecasts for the 16 WNP TCs in 2023 (Supplementary Fig. S13). DaYu-Global achieves the lowest RMSE among the evaluated global AI models from 24 h onwards, with errors remaining near 24 m $s^{-1}$ at 72–120 h (Fig. 2f). At 120 h, its RMSE is approximately 22%, 8%, 10% and 6% lower than those of FengWu, FuXi, GraphCast and Pangu-Weather, respectively. This sustained intensity advantage is accompanied by track accuracy broadly comparable to that of FengWu, GraphCast and Pangu-Weather over the first 72 h (Fig. 2e). Together, these results show that DaYu-Global improves TC intensity prediction while retaining competitive track forecast skill.

Storm-specific evaluations further show that DaYu-Global's intensity advantage spans TCs with different peak intensities (Supplementary Fig. S3). Wind-field and track comparisons for Koinu, Doksuri and Bolaven show that DaYu-Global also preserves concentrated near-centre strong winds during multi-step prediction (Supplementary Figs. S4–S6).

### 3.2. Regional gains in TC intensity prediction

DaYu-TC couples DaYu-Global with the high-resolution DaYu-Region model through one-way nesting. The resulting system consistently improves TC intensity forecasts in terms of both minimum pressure and maximum wind speed (Fig. 3). Across 12–120 h, minimum-pressure RMSE decreases by 28.6–39.8% and maximum-wind RMSE by 22.0–33.1% relative to DaYu-Global. At 120 h, the respective reductions remain 28.6% and 24.4%, demonstrating sustained regional gains throughout the five-day forecast.

Compared with concurrent operational IFS-HRES forecasts, DaYu-TC achieves lower minimum-pressure RMSE at every lead time shown and a clear maximum-wind advantage at 48–96 h. At 72 h, its minimum-pressure and maximum-wind RMSEs are 17.2 hPa and 16.6 m $s^{-1}$, respectively, approximately 26% and 12% lower than those of IFS-HRES. These results show that combining global extreme preservation with regional dynamic prediction improves both key measures of TC intensity, enabling DaYu-TC to outperform IFS-HRES for the indicated lead times.

Storm-specific comparisons show that the regional gains extend across individual TCs (Fig. 4). DaYu-TC reduces maximum-wind RMSE relative to DaYu-Global for all 16 storms and minimum-pressure RMSE for 15. IFS-HRES remains a strong benchmark, achieving the lowest intensity errors for several storms, while DaYu-TC achieves lower maximum-wind and minimum-pressure RMSE for 9 and 12 storms, respectively. Relative to IFS-HRES, the gains are particularly pronounced for Haikui and Bolaven: maximum-wind RMSE decreases by 38.4% and 20.3%, respectively, and minimum-pressure RMSE by 46.1% and 30.1%. For Mawar, minimum-pressure RMSE decreases from 24.1 to 17.0 hPa. These results demonstrate that the aggregate intensity gains extend across storms with a wide range of peak intensities.

To examine the structural changes associated with regional intensity gains, we analyse forecasts of TC Koinu initialized at 12:00 UTC on 30 September 2023 (Fig. 5a–t). On days 2–3, DaYu-TC produces a more concentrated ring of strong winds and sharper radial wind-speed gradients than DaYu-Global.

Absolute maximum-wind and minimum-pressure biases are reduced at every lead time shown. On day 3, the negative maximum-wind bias is reduced from −21.5 to −3.5 m $s^{-1}$ (Fig. 5k–l). Compared with IFS-HRES, DaYu-TC has smaller absolute minimum-pressure biases at every lead time shown and smaller absolute maximum-wind biases on days 1–4. Its day-5 centre displacement is 126 km, compared with 212 km for DaYu-Global and 206 km for IFS-HRES.

Notably, the high-resolution IFS analysis underestimates Koinu's intensity, with negative maximum-wind biases and positive minimum-pressure biases at every valid time shown. This pattern is consistent with documented limitations of reanalyses and operational analyses in representing TC inner-core winds and pressure, associated with effective resolution, observational coverage and data assimilation [20,21,38]. On day 5, DaYu-TC reduces the maximum-wind bias from −13.8 m $s^{-1}$ in the analysis to −4.9 m $s^{-1}$ and the minimum-pressure bias from +24.3 to −0.3 hPa (Fig. 5q, t). The regional-nesting gains in this case therefore include both a stronger near-centre wind field and more accurate intensity relative to best-track estimates.

Additional Doksuri and Bolaven cases also show more concentrated near-centre strong-wind regions after regional nesting, providing further spatial evidence for the intensity gains (Supplementary Figs. S7 and S8). Beyond TC forecasts, DaYu-TC improves conventional regional forecasts, reducing the 6 h RMSE of all nine variables by approximately 17.5–44.6% (Supplementary Fig. S9). Improvements in both RMSE and ACC persist throughout 120 h for 2 m temperature (T2M), MSLP and 850 hPa temperature (T850).

### 3.3. Regional gains across global drivers

The same DaYu-Region weights are evaluated under four global drivers: DaYu-Global, FengWu, Pangu-Weather and GraphCast (Fig. 6). The DaYu-Global-driven configuration serves as the reference, while the other three configurations assess whether the regional model retains its forecast gains under different global drivers without retraining or fine-tuning.

Regional nesting reduces both minimum-pressure and maximum-wind RMSE at every lead time from 12 to 120 h under all four drivers. For the default DaYu-TC configuration, overall minimum-pressure and maximum-wind RMSE decrease by 35.7% and 28.5%, respectively. Under FengWu, Pangu-Weather and GraphCast, the corresponding pressure reductions are 29.4%, 33.2% and 28.8%. The wind-speed reductions are 22.6%, 24.1% and 28.6%. Positive gains persist to 120 h for every driver, demonstrating that the regional model improves TC intensity forecasts across distinct global backgrounds with a single set of model weights.

The Doksuri, Koinu and Bolaven cases illustrate how these cross-driver intensity gains are expressed spatially, with stronger and more concentrated near-centre wind fields in the regional forecasts (Supplementary Figs. S10–S12).

### 3.4. Computational efficiency

DaYu-TC generates a 120 h forecast in 47.3 s on a single NVIDIA A100 GPU, with peak allocated GPU memory of 16.0 GiB (Supplementary Table S1). Separate benchmarks quantify the costs of DaYu-Global and DaYu-Region. The IFS-HRES runtime reported by Lam et al. [6] is provided as context for a 240 h global forecast.

## 4. Discussion and conclusion

Persistent intensity underestimation remains a major challenge for AI tropical cyclone forecasting. DaYu-TC (Fig. 1) addresses this challenge by combining global extreme preservation with dynamic regional refinement through a shared architecture and extreme-aware training objectives. In the 2023 western North Pacific evaluation, it achieves lower minimum-pressure RMSE than concurrent IFS-HRES forecasts throughout 12–120 h and lower maximum-wind RMSE at 48–96 h, with respective reductions of 26% and 12% at 72 h. The same regional weights improve intensity forecasts driven by three additional global AI models without retraining. These results demonstrate a transferable multiscale approach that advances TC intensity prediction across global forecasting systems.

The global and regional evaluations reveal complementary opportunities for improving TC intensity prediction. At the global scale, DaYu-Global reduces the attenuation of strong-wind and low-pressure anomalies while retaining competitive conventional forecast skill (Fig. 2a–d; Supplementary Fig. S1). Its consistently lower maximum-wind RMSE than the other evaluated global AI models supports the value of preserving extreme signals for TC intensity prediction (Fig. 2f). Regional nesting combines local evolution with the forecast large-scale environment, reducing wind and pressure errors and producing more concentrated near-centre strong winds and sharper gradients (Figs. 3 and 5). These results highlight the complementary benefits of global extreme preservation and regional refinement for TC intensity forecasting.

Regional gains extend across TCs with different peak intensities (Fig. 4). DaYu-TC reduces maximum-wind and minimum-pressure RMSE relative to DaYu-Global for 16 and 15 storms, respectively, and outperforms IFS-HRES for 9 and 12 storms. The Koinu case further illustrates that these gains can include reduced intensity underestimation relative to the verifying analysis. Although operational analyses can underestimate TC intensity owing to their representation of inner-core structures [20,21,38], DaYu-TC produces intensity estimates closer to best-track records at some lead times (Fig. 5). Initialization from DaYu-Global may help DaYu-Region draw on TC evolution learned across different regions. Sun et al. [13] demonstrated that AI models can learn from intense TCs in one basin to predict intense TCs in another, even when such storms are absent from the target basin's training data. In DaYu-TC, this potential transfer of information may contribute to the improved intensity estimates.

The cross-driver experiments extend the contribution from a particular global–regional combination to a reusable regional forecasting component. DaYu-Region was trained with DaYu-Global backgrounds, yet the same model weights consistently reduce maximum-wind and minimum-pressure RMSE when driven by Pangu-Weather, GraphCast and FengWu (Fig. 6). The regional gains therefore persist despite differences among the global forecasts, supporting the use of a shared regional model to enhance multiple AI systems. DaYu-TC completes a 120 h forecast in under one minute on a single NVIDIA A100 GPU (Supplementary Table S1). This combination of transferability and low inference cost provides a practical basis for regional ensembles driven by different global forecasts [8,39].

Further research could examine how uncertainty in the global environment propagates into regional TC intensity and rapid intensification. Overall, DaYu-TC combines global extreme preservation with dynamic regional prediction to reduce TC intensity underestimation and improve wind and pressure forecasts. Its ability to enhance multiple global AI models without retraining offers an efficient, reusable approach to TC intensity forecasting.

## Conflict of interest

None declared.

## Acknowledgments

The authors thank ECMWF for providing ERA5 reanalysis and IFS-HRES analysis and forecasts and NOAA for providing the IBTrACS TC best-track data. We thank the Pangu-Weather, GraphCast, FuXi and FengWu development teams for making their models and associated resources available. We also thank the Shanghai Artificial Intelligence Laboratory for computational resources and technical support.

## Funding

This work was supported by the National Natural Science Foundation of China (42450254 and 42575069), the Meteorological Joint Funds of the National Natural Science Foundation of China (U2542213) and the AI for Math and Science programme of the Shanghai Institute for Mathematics and Interdisciplinary Sciences (SIMS-ID-26-AMS-004).

## Author contributions

Runzepu Hou and Feng Zhang conceived and designed the study and drafted the manuscript. Xin Wu and Xujun Wei conducted the tropical cyclone forecast evaluation and analysed the results. Fenghua Ling provided computational resources and contributed to data analysis. Kan Dai, Gaozhen Nie, Leiming Ma and Renhe Zhang analysed the results and contributed to writing the manuscript. All authors participated in the research and discussions, revised the manuscript and approved the final version.

## References

[1] Young R, Hsiang S. Mortality caused by tropical cyclones in the United States. Nature 2024; 635: 121–8. doi: 10.1038/s41586-024-07945-5.

[2] Shi R, Min J, Dong C, et al. Rising coastal wave hazards from climate-driven winds and tropical cyclones. Natl Sci Rev 2026; 13: nwag528. doi: 10.1093/nsr/nwag528.

[3] Cangialosi JP, Blake E, DeMaria M, et al. Recent progress in tropical cyclone intensity forecasting at the National Hurricane Center. Weather Forecast 2020; 35: 1913–22. doi: 10.1175/WAF-D-20-0059.1.

[4] Yu H, Chen G, Wong WK, et al. WMO Typhoon Landfall Forecast Demonstration Project (2010–22): A Decade of Transition from Track Forecasts to Impact Forecasts. Bull Am Meteorol Soc 2024; 105: E1320–49. doi: 10.1175/BAMS-D-23-0085.1.

[5] Bi K, Xie L, Zhang H, et al. Accurate medium-range global weather forecasting with 3D neural networks. Nature 2023; 619: 533–8. doi: 10.1038/s41586-023-06185-3.

[6] Lam R, Sanchez-Gonzalez A, Willson M, et al. Learning skillful medium-range global weather forecasting. Science 2023; 382: 1416–21. doi: 10.1126/science.adi2336.

[7] Chen L, Zhong X, Zhang F, et al. FuXi: A cascade machine learning forecasting system for 15-day global weather forecast. npj Clim Atmos Sci 2023; 6: 190. doi: 10.1038/s41612-023-00512-1.

[8] Luo JJ, Xia J, Pan B, et al. AI for atmosphere–ocean sciences: advancements, challenges and ways forward. Natl Sci Rev 2026; 13: nwag063. doi: 10.1093/nsr/nwag063.

[9] Olivetti L, Messori G. Do data-driven models beat numerical models in forecasting weather extremes? A comparison of IFS HRES, Pangu-Weather and GraphCast. Geosci Model Dev 2024; 17: 7915–62. doi: 10.5194/gmd-17-7915-2024.

[10] Sahu PL, Sandeep S, Kodamana H. Evaluating global machine learning models for tropical cyclone dynamics and thermodynamics. J Geophys Res Mach Learn Comput 2025; 2: e2025JH000594. doi: 10.1029/2025JH000594.

[11] Shi Y, Hu R, Wu N, et al. Comparison of AI and NWP models in operational severe weather forecasting: A study on tropical cyclone predictions. J Geophys Res Mach Learn Comput 2025; 2: e2024JH000481. doi: 10.1029/2024JH000481.

[12] DeMaria M, Franklin JL, Chirokova G, et al. An operations-based evaluation of tropical cyclone track and intensity forecasts from artificial intelligence weather prediction models. Artif Intell Earth Syst 2025; 4: 240085. doi: 10.1175/AIES-D-24-0085.1.

[13] Sun YQ, Hassanzadeh P, Zand M, et al. Can AI weather models predict out-of-distribution gray swan tropical cyclones? Proc Natl Acad Sci USA 2025; 122: e2420914122. doi: 10.1073/pnas.2420914122.

[14] Subich C, Husain SZ, Separovic L, et al. Fixing the double penalty in data-driven weather forecasting through a modified spherical harmonic loss function. In: Proceedings of the 42nd International Conference on Machine Learning. Proc Mach Learn Res 2025; 267: 57191–211. https://proceedings.mlr.press/v267/subich25a.html

[15] Bonavita M. On Some Limitations of Current Machine Learning Weather Prediction Models. Geophys Res Lett 2024; 51: e2023GL107377. doi: 10.1029/2023GL107377.

[16] Lopez-Gomez I, McGovern A, Agrawal S, et al. Global Extreme Heat Forecasting Using Neural Weather Models. Artif Intell Earth Syst 2023; 2: e220035. doi: 10.1175/AIES-D-22-0035.1.

[17] Zhong X, Chen L, Liu J, et al. FuXi-Extreme: Improving extreme rainfall and wind forecasts with diffusion model. Sci China Earth Sci 2024; 67: 3696–708. doi: 10.1007/s11430-023-1427-x.

[18] An S, Kim Y, Cho D, et al. Toward spatially sharper precipitation prediction via global-local frequency guidance. npj Clim Atmos Sci 2026; 9: 200. doi: 10.1038/s41612-026-01472-y.

[19] Qin C, Ding Y, Tao P, et al. Forecasting extreme temperature events by Hankel-augmented contrastive learning. Natl Sci Rev 2026; 13: nwag456. doi: 10.1093/nsr/nwag456.

[20] Hodges K, Cobb A, Vidale PL. How Well Are Tropical Cyclones Represented in Reanalysis Datasets? J Clim 2017; 30: 5243–64. doi: 10.1175/JCLI-D-16-0557.1.

[21] Liu G, Jiang S, Zheng M, et al. A Global ERA5-based Tropical Cyclone Wind Field Dataset Enhanced by Integrated Parametric Correction Methods. Sci Data 2025; 12: 1429. doi: 10.1038/s41597-025-05789-w.

[22] Skamarock WC, Klemp JB, Dudhia J, et al. A description of the Advanced Research WRF version 3. NCAR Technical Note NCAR/TN-475+STR. Boulder, CO: National Center for Atmospheric Research, 2008. doi: 10.5065/D68S4MVH.

[23] Tallapragada V, Kieu C, Trahan S, et al. Forecasting tropical cyclones in the Western North Pacific Basin using the NCEP operational HWRF Model: Model upgrades and evaluation of real-time performance in 2013. Weather Forecast 2016; 31: 877–94. doi: 10.1175/WAF-D-14-00139.1.

[24] Pathak J, Cohen Y, Garg P, et al. Kilometer-scale convection-allowing model emulation using generative diffusion modeling. Sci Adv 2026; 12: eadv0423. doi: 10.1126/sciadv.adv0423.

[25] Gao Y, Wu H, Shu R, et al. OneForecast: a universal framework for global and regional weather forecasting. In: Proceedings of the 42nd International Conference on Machine Learning. Proc Mach Learn Res 2025; 267: 18658–97. https://proceedings.mlr.press/v267/gao25r.html

[26] Nipen TN, Haugen HH, Ingstad MS, et al. Regional Data-Driven Weather Modeling with a Global Stretched Grid. Artif Intell Earth Syst 2026; 5: 250001. doi: 10.1175/AIES-D-25-0001.1.

[27] Baño-Medina J, Sengupta A, Steinhoff D, et al. A regional high resolution AI weather model for the prediction of atmospheric rivers and extreme precipitation. npj Clim Atmos Sci 2025; 8: 385. doi: 10.1038/s41612-025-01265-9.

[28] Guo S, Chen L, Zhao Y, et al. FuXi-TC: a generative framework integrating deep learning and physics-based models for improved tropical cyclone forecasts. npj Clim Atmos Sci 2026. doi: 10.1038/s41612-026-01501-w.

[29] Niu Z, Huang W, Huang S, et al. A regional artificial intelligence model for skillful typhoon prediction. npj Nat Hazards 2026; 3: 61. doi: 10.1038/s44304-026-00219-2.

[30] Hersbach H, Bell B, Berrisford P, et al. The ERA5 global reanalysis. Q J R Meteorol Soc 2020; 146: 1999–2049. doi: 10.1002/qj.3803.

[31] ECMWF. IFS Documentation CY47R3—Part II: Data assimilation. Reading: ECMWF, 2021. doi: 10.21957/t445u8kna.

[32] Knapp KR, Kruk MC, Levinson DH, et al. The International Best Track Archive for Climate Stewardship (IBTrACS): Unifying tropical cyclone best track data. Bull Am Meteorol Soc 2010; 91: 363–76. doi: 10.1175/2009BAMS2755.1.

[33] Gahtan J, Knapp KR, Schreck CJ, et al. International Best Track Archive for Climate Stewardship (IBTrACS) Project, Version 4r01. NOAA National Centers for Environmental Information, 2024. doi: 10.25921/82ty-9e16.

[34] Loshchilov I, Hutter F. Decoupled weight decay regularization. In: Proceedings of the International Conference on Learning Representations, New Orleans, LA, 6–9 May 2019. https://openreview.net/forum?id=Bkg6RiCqY7

[35] Chen K, Han T, Ling F, et al. The operational medium-range deterministic weather forecasting can be extended beyond a 10-day lead time. Commun Earth Environ 2025; 6: 518. doi: 10.1038/s43247-025-02502-y.

[36] Rasp S, Hoyer S, Merose A, et al. WeatherBench 2: A benchmark for the next generation of data-driven global weather models. J Adv Model Earth Syst 2024; 16: e2023MS004019. doi: 10.1029/2023MS004019.

[37] Zhang X, Hegerl G, Zwiers FW, et al. Avoiding inhomogeneity in percentile-based indices of temperature extremes. J Clim 2005; 18: 1641–51. doi: 10.1175/JCLI3366.1.

[38] Magnusson L, Majumdar S, Emerton R, et al. Tropical cyclone activities at ECMWF. ECMWF Technical Memorandum 888. Reading: ECMWF, 2021. doi: 10.21957/zzxzzygwv.

[39] Price I, Sanchez-Gonzalez A, Alet F, et al. Probabilistic weather forecasting with machine learning. Nature 2025; 637: 84–90. doi: 10.1038/s41586-024-08252-9.

# Figure captions

Figure 1. Global–regional nesting framework of DaYu-TC. a, DaYu-Global uses two consecutive 0.25° global atmospheric states to predict the state 6 h ahead. b, the target region is extracted from the global forecast and interpolated from 0.25° to 0.1° to provide the background valid at the regional prediction time. c, DaYu-Region combines two consecutive 0.1° regional states with this background to predict the next regional atmospheric state. During training, regional states and targets are taken from IFS analyses; during autoregressive inference, predicted states are fed back as inputs.

Figure 2. Global extreme-anomaly preservation and TC forecast performance of DaYu-Global. a, relative mean bias error (rMBE) at 120 h on the 2023 test set for 10 m zonal and meridional winds (U10 and V10), 850 hPa zonal and meridional winds (U850 and V850) and mean sea-level pressure (MSLP). Red crosses and labels indicate DaYu-Global; the dashed line denotes zero bias. b–d, reductions in absolute rMBE relative to each comparator at 24, 72 and 120 h, in percentage points; positive values favour DaYu-Global. e, mean Euclidean track error (km). f, maximum 10 m wind-speed RMSE (m $s^{-1}$). Panels e and f evaluate 16 WNP TCs in 2023 against NOAA IBTrACS best-track estimates at 12–120 h. Parentheses below the axes give valid sample counts. All panels compare DaYu-Global with FengWu, FuXi, GraphCast and Pangu-Weather.

Figure 3. Improvements in TC pressure and wind-speed forecasts by DaYu-TC. a, minimum mean sea-level pressure RMSE (hPa). b, maximum 10 m wind-speed RMSE (m $s^{-1}$). DaYu-Global, DaYu-TC and IFS-HRES forecasts are evaluated against NOAA IBTrACS best-track estimates for 16 TCs in the WNP in 2023 at lead times of 12–120 h. Grey horizontal lines connect DaYu-Global and DaYu-TC errors at each lead time; percentages on the right indicate RMSE reductions relative to DaYu-Global. Numbers in parentheses on the vertical axes indicate valid sample counts. Smaller horizontal-axis values indicate lower errors.

Figure 4. Storm-specific TC intensity forecast performance of DaYu-TC. Maximum 10 m wind-speed RMSE (m $s^{-1}$) and minimum mean sea-level pressure RMSE (hPa) are compared for DaYu-Global, IFS-HRES and DaYu-TC against NOAA IBTrACS best-track estimates for 16 TCs in the WNP in 2023. For each storm, errors are pooled over matched samples at 12–120 h. Storms are grouped by peak intensity and ordered by increasing best-track peak wind speed, shown in parentheses. Cell values give RMSE; colours indicate within-storm model ranks, calculated separately for wind and pressure, with rank 1 denoting the lowest RMSE. Horizontal bars show the percentage RMSE reduction relative to IFS-HRES. Teal and red bars indicate positive and negative reductions, respectively.

Figure 5. Near-centre wind and pressure structures of TC Koinu predicted by DaYu-TC. Forecasts initialized at 12:00 UTC on 30 September 2023 are shown at lead times of 1–5 d alongside analyses at the corresponding valid times. Columns show the operational IFS analysis, IFS-HRES, DaYu-Global and DaYu-TC, from left to right; rows correspond to days 1–5. Shading indicates 10 m wind speed (m $s^{-1}$) and grey contours show mean sea-level pressure (hPa). Each panel is centred on its own diagnosed TC centre; horizontal and vertical axes indicate eastward and northward distances (km), respectively. ΔVmax and ΔPmin denote biases in maximum 10 m wind speed (m $s^{-1}$) and minimum mean sea-level pressure (hPa), respectively. Biases are calculated as the value diagnosed from the forecast or analysis minus the NOAA best-track estimate at the same valid time. *D*pos in the lower-left corner denotes the distance between the diagnosed TC centre and the NOAA best-track centre (km).

Figure 6. TC intensity gains from regional nesting under four global forecast drivers. Grouped bars show percentage RMSE reductions for a, minimum mean sea-level pressure and b, maximum 10 m wind speed at 12–120 h for 16 TCs in the WNP in 2023. Colours identify DaYu-Global, FengWu, Pangu-Weather and GraphCast. All regional forecasts use the same DaYu-Region weights trained with DaYu-Global backgrounds. Errors are evaluated against NOAA IBTrACS best-track estimates. Reductions are calculated relative to each global driver as $100 \times (\mathrm{RMSE}_{global} - \mathrm{RMSE}_{regional})/\mathrm{RMSE}_{global}$. Positive values indicate lower RMSE after nesting. Numbers in parentheses below the horizontal axes give valid sample counts. Percentages below each panel report reductions in overall RMSE, obtained by pooling squared errors across lead times before taking the square root.

**Fig. 1**

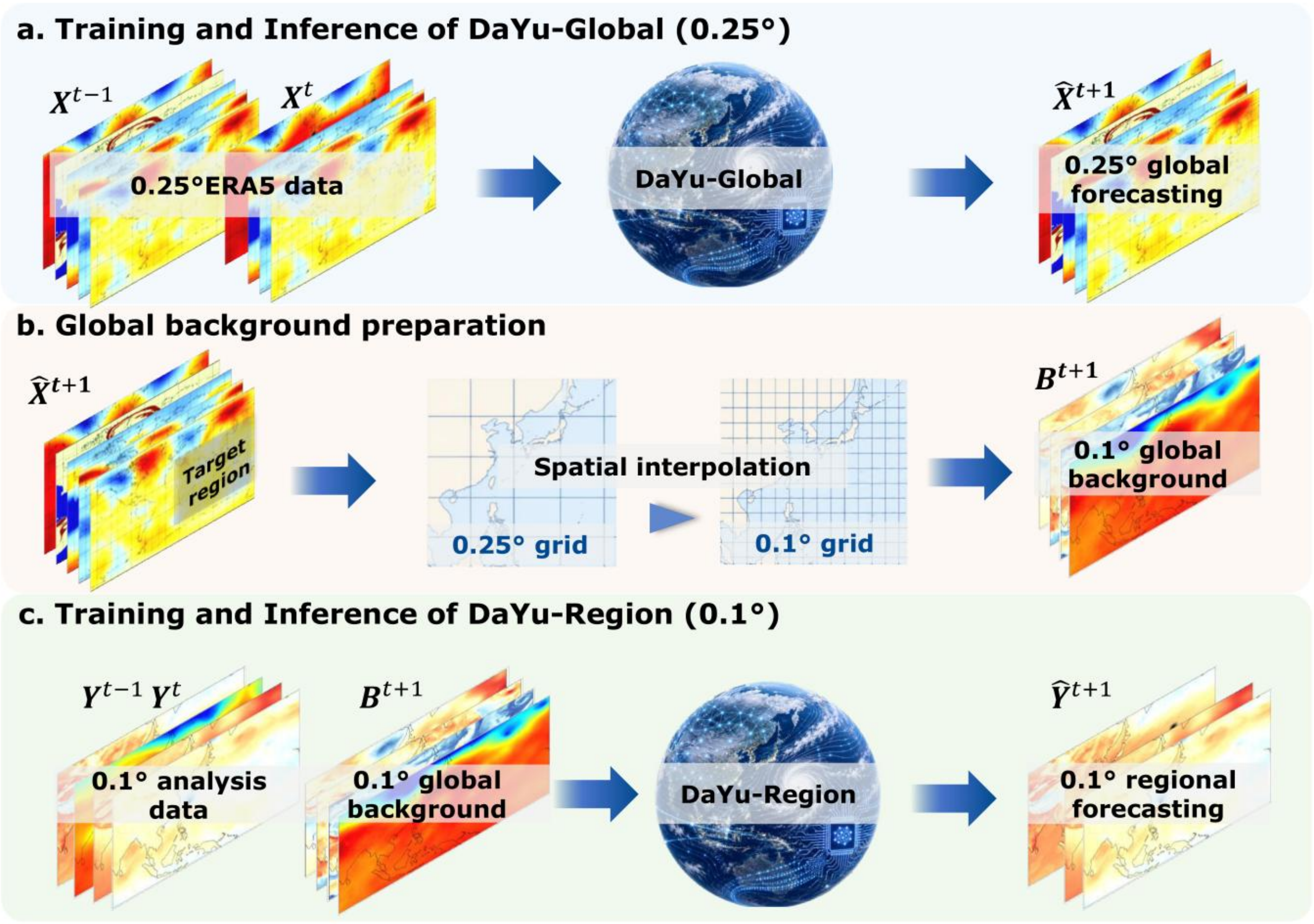

a. Training and Inference of DaYu-Global (0.25°)
$X^{t-1}$
$X^{t}$
0.25°ERA5 data
DaYu-Global
$\hat{X}^{t+1}$
0.25° global forecasting
b. Global background preparation
$\hat{X}^{t+1}$
Target region
Spatial interpolation
0.25° grid
0.1° grid
$B^{t+1}$
0.1° global background
c. Training and Inference of DaYu-Region (0.1°)
$Y^{t-1}$ $Y^{t}$
0.1° analysis data
$B^{t+1}$
0.1° global background
DaYu-Region
$\hat{Y}^{t+1}$
0.1° regional forecasting

Fig. 2

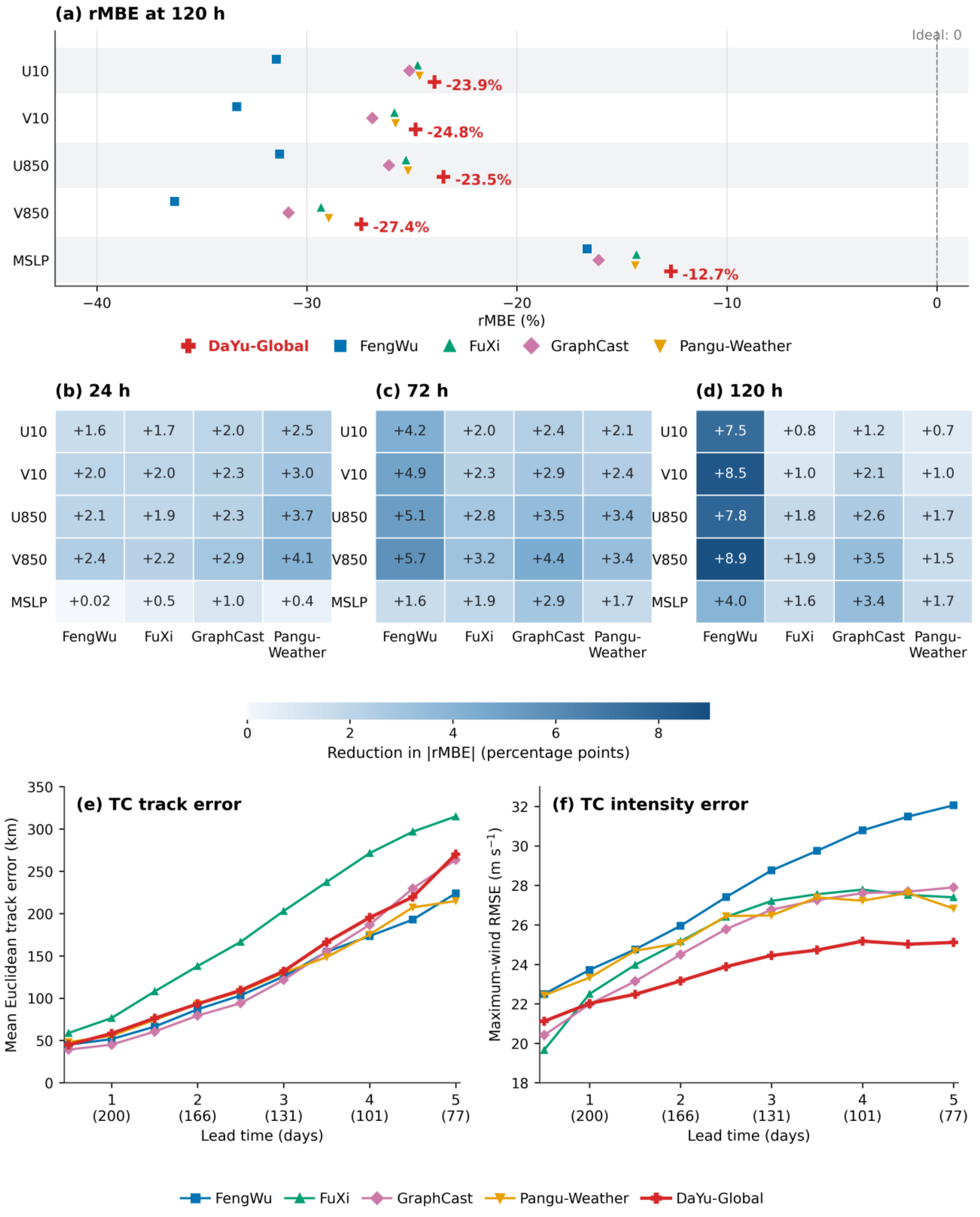

(a) rMBE at 120 h
Ideal: 0
U10
V10
U850
V850
MSLP
-23.9%
-24.8%
-23.5%
-27.4%
-12.7%
−40
−30
−20
−10
0
rMBE (%)
DaYu-Global
FengWu
FuXi
GraphCast
Pangu-Weather
(b) 24 h
(c) 72 h
(d) 120 h
U10 +1.6 +1.7 +2.0 +2.5
V10 +2.0 +2.0 +2.3 +3.0
U850 +2.1 +1.9 +2.3 +3.7
V850 +2.4 +2.2 +2.9 +4.1
MSLP +0.02 +0.5 +1.0 +0.4
U10 +4.2 +2.0 +2.4 +2.1
V10 +4.9 +2.3 +2.9 +2.4
U850 +5.1 +2.8 +3.5 +3.4
V850 +5.7 +3.2 +4.4 +3.4
MSLP +1.6 +1.9 +2.9 +1.7
U10 +7.5 +0.8 +1.2 +0.7
V10 +8.5 +1.0 +2.1 +1.0
U850 +7.8 +1.8 +2.6 +1.7
V850 +8.9 +1.9 +3.5 +1.5
MSLP +4.0 +1.6 +3.4 +1.7
FengWu
FuXi
GraphCast
Pangu-Weather
0
2
4
6
8
Reduction in |rMBE| (percentage points)
(e) TC track error
Mean Euclidean track error (km)
0
50
100
150
200
250
300
350
1 (200)
2 (166)
3 (131)
4 (101)
5 (77)
Lead time (days)
(f) TC intensity error
Maximum-wind RMSE (m s−1)
18
20
22
24
26
28
30
32
1 (200)
2 (166)
3 (131)
4 (101)
5 (77)
Lead time (days)
FengWu
FuXi
GraphCast
Pangu-Weather
DaYu-Global

**Fig. 3**

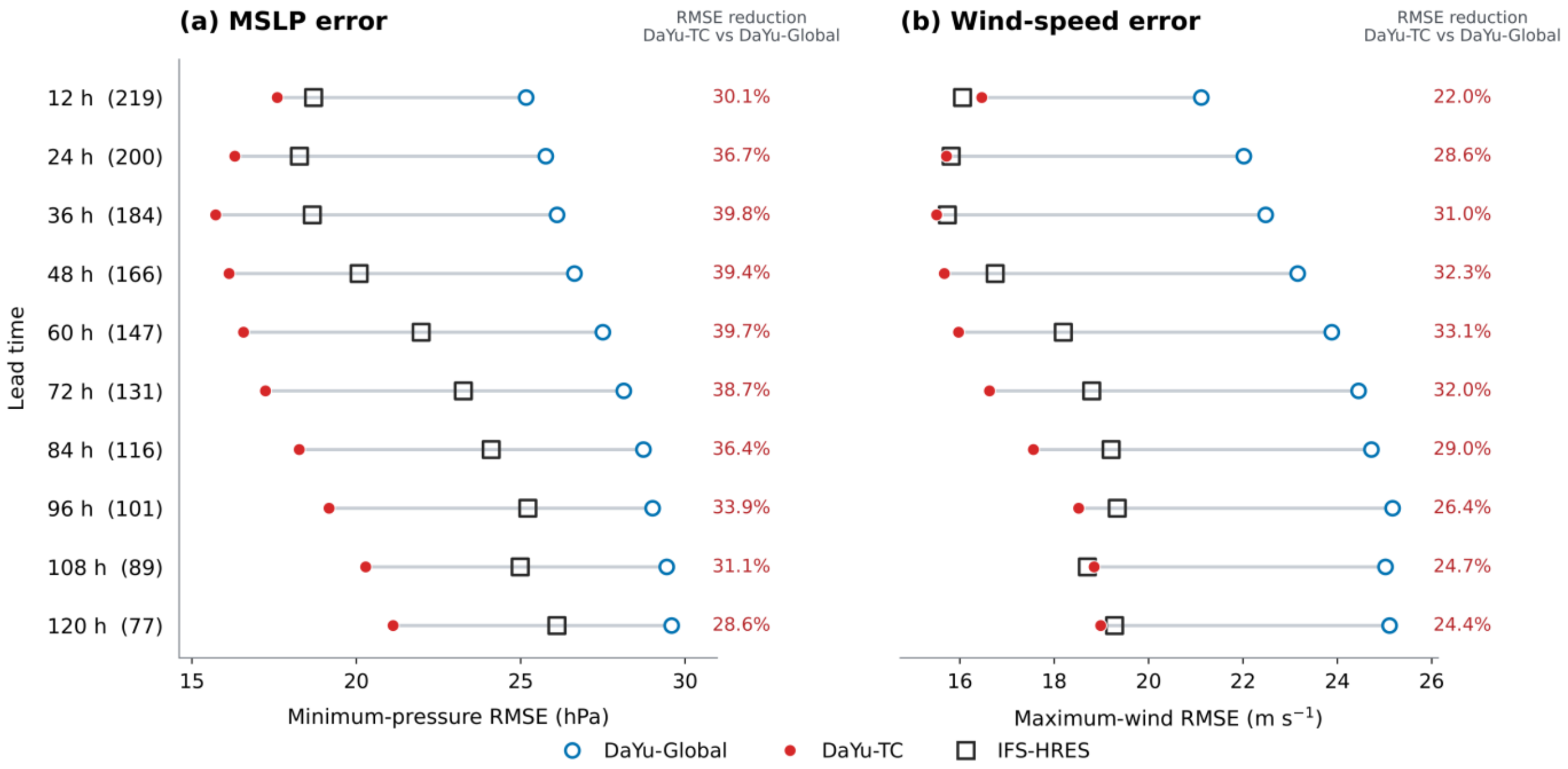

(a) MSLP error
RMSE reduction DaYu-TC vs DaYu-Global
(b) Wind-speed error
RMSE reduction DaYu-TC vs DaYu-Global
Lead time
12 h (219)
24 h (200)
36 h (184)
48 h (166)
60 h (147)
72 h (131)
84 h (116)
96 h (101)
108 h (89)
120 h (77)
30.1%
36.7%
39.8%
39.4%
39.7%
38.7%
36.4%
33.9%
31.1%
28.6%
22.0%
28.6%
31.0%
32.3%
33.1%
32.0%
29.0%
26.4%
24.7%
24.4%
15
20
25
30
Minimum-pressure RMSE (hPa)
16
18
20
22
24
26
Maximum-wind RMSE (m s$^{-1}$)
DaYu-Global
DaYu-TC
IFS-HRES

**Fig. 4**

| Tropical cyclone (peak wind, m s$^{-1}$) | Maximum-wind RMSE (m s$^{-1}$) DaYu-Global | Maximum-wind RMSE (m s$^{-1}$) IFS-HRES | Maximum-wind RMSE (m s$^{-1}$) **DaYu-TC** | Minimum-pressure RMSE (hPa) DaYu-Global | Minimum-pressure RMSE (hPa) IFS-HRES | Minimum-pressure RMSE (hPa) **DaYu-TC** | DaYu-TC RMSE reduction vs IFS-HRES (%) Wind | DaYu-TC RMSE reduction vs IFS-HRES (%) Pressure |
|---|---|---|---|---|---|---|---|---|
| **Tropical storms** | | | | | | | | |
| Sanba (20.1) | 2.9 | 3.1 | **2.6** | 3.8 | **3.6** | 4.3 | +15.4% | -20.5% |
| Yun-Yeung (20.1) | 3.1 | 2.3 | **2.1** | 6.8 | 5.8 | **5.1** | +10.6% | +11.6% |
| Sanvu (23.1) | 8.5 | **5.3** | 5.5 | 7.3 | **4.3** | 4.4 | -3.8% | -1.5% |
| Kirogi (30.9) | 5.1 | 3.9 | **3.7** | 6.5 | **3.7** | 4.7 | +4.1% | -25.7% |
| Damrey (32.9) | 6.6 | 6.9 | **5.4** | 12.6 | 12.3 | **10.6** | +21.9% | +14.5% |
| **Typhoons** | | | | | | | | |
| Talim (36.0) | 11.1 | **7.5** | 9.9 | 6.0 | 6.1 | **4.9** | -31.0% | +18.4% |
| Guchol (45.8) | 15.6 | 14.3 | **12.1** | 14.6 | 15.3 | **11.2** | +15.0% | +26.8% |
| Dora (49.9) | 3.9 | **2.9** | 3.6 | 13.1 | 10.9 | **10.1** | -24.3% | +7.4% |
| Haikui (54.0) | 18.8 | 18.7 | **11.5** | 20.2 | 22.7 | **12.2** | +38.4% | +46.1% |
| Lan (59.2) | 15.3 | 12.5 | **11.1** | 14.0 | 12.9 | **10.5** | +10.8% | +18.0% |
| Khanun (61.2) | 13.2 | 12.0 | **11.2** | 13.1 | 15.7 | **10.4** | +6.6% | +33.9% |
| Koinu (61.2) | 25.3 | **17.4** | 18.3 | 31.4 | **21.6** | 23.5 | -4.9% | -8.8% |
| Doksuri (65.8) | 27.6 | **18.4** | 20.4 | 32.8 | 23.2 | **22.1** | -10.9% | +4.7% |
| **Super typhoons** | | | | | | | | |
| Saola (68.9) | 35.2 | **20.4** | 23.3 | 44.1 | 23.9 | **23.6** | -14.4% | +1.1% |
| Bolaven (79.2) | 35.6 | 34.6 | **27.5** | 44.2 | 43.6 | **30.5** | +20.3% | +30.1% |
| Mawar (81.8) | 27.5 | **19.2** | 19.3 | 28.7 | 24.1 | **17.0** | -0.1% | +29.3% |

Bar chart axes (Wind, Pressure): −80, 0, 80

Within-storm rank: 1 (lowest RMSE) 2 3

**Fig. 5**

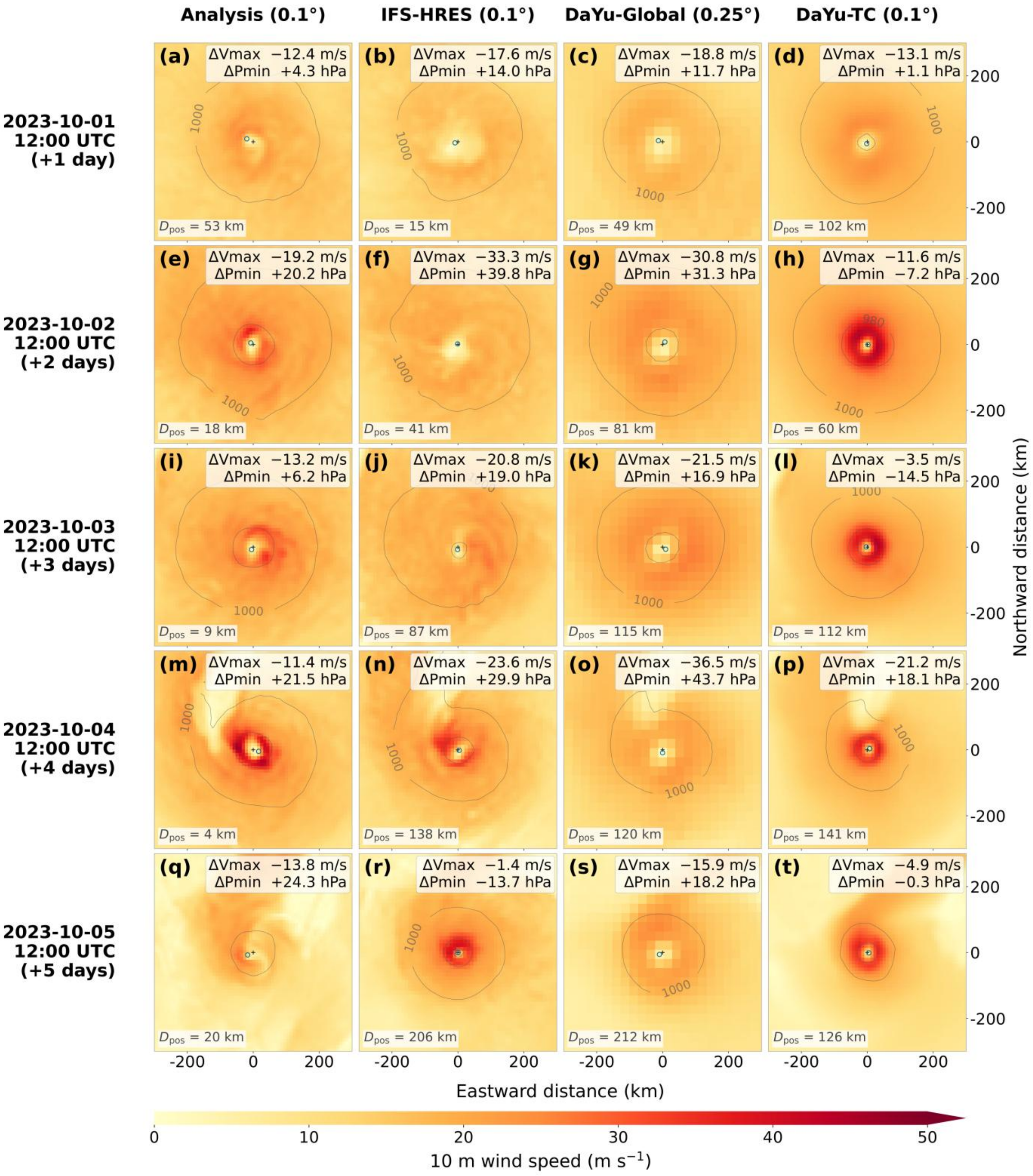

Analysis (0.1°)
IFS-HRES (0.1°)
DaYu-Global (0.25°)
DaYu-TC (0.1°)
2023-10-01 12:00 UTC (+1 day)
2023-10-02 12:00 UTC (+2 days)
2023-10-03 12:00 UTC (+3 days)
2023-10-04 12:00 UTC (+4 days)
2023-10-05 12:00 UTC (+5 days)
(a) ΔVmax −12.4 m/s ΔPmin +4.3 hPa
$D_{pos}$ = 53 km
(b) ΔVmax −17.6 m/s ΔPmin +14.0 hPa
$D_{pos}$ = 15 km
(c) ΔVmax −18.8 m/s ΔPmin +11.7 hPa
$D_{pos}$ = 49 km
(d) ΔVmax −13.1 m/s ΔPmin +1.1 hPa
$D_{pos}$ = 102 km
(e) ΔVmax −19.2 m/s ΔPmin +20.2 hPa
$D_{pos}$ = 18 km
(f) ΔVmax −33.3 m/s ΔPmin +39.8 hPa
$D_{pos}$ = 41 km
(g) ΔVmax −30.8 m/s ΔPmin +31.3 hPa
$D_{pos}$ = 81 km
(h) ΔVmax −11.6 m/s ΔPmin −7.2 hPa
$D_{pos}$ = 60 km
(i) ΔVmax −13.2 m/s ΔPmin +6.2 hPa
$D_{pos}$ = 9 km
(j) ΔVmax −20.8 m/s ΔPmin +19.0 hPa
$D_{pos}$ = 87 km
(k) ΔVmax −21.5 m/s ΔPmin +16.9 hPa
$D_{pos}$ = 115 km
(l) ΔVmax −3.5 m/s ΔPmin −14.5 hPa
$D_{pos}$ = 112 km
(m) ΔVmax −11.4 m/s ΔPmin +21.5 hPa
$D_{pos}$ = 4 km
(n) ΔVmax −23.6 m/s ΔPmin +29.9 hPa
$D_{pos}$ = 138 km
(o) ΔVmax −36.5 m/s ΔPmin +43.7 hPa
$D_{pos}$ = 120 km
(p) ΔVmax −21.2 m/s ΔPmin +18.1 hPa
$D_{pos}$ = 141 km
(q) ΔVmax −13.8 m/s ΔPmin +24.3 hPa
$D_{pos}$ = 20 km
(r) ΔVmax −1.4 m/s ΔPmin −13.7 hPa
$D_{pos}$ = 206 km
(s) ΔVmax −15.9 m/s ΔPmin +18.2 hPa
$D_{pos}$ = 212 km
(t) ΔVmax −4.9 m/s ΔPmin −0.3 hPa
$D_{pos}$ = 126 km
1000
980
-200
0
200
Eastward distance (km)
Northward distance (km)
0
10
20
30
40
50
10 m wind speed (m s⁻¹)

**Fig. 6**

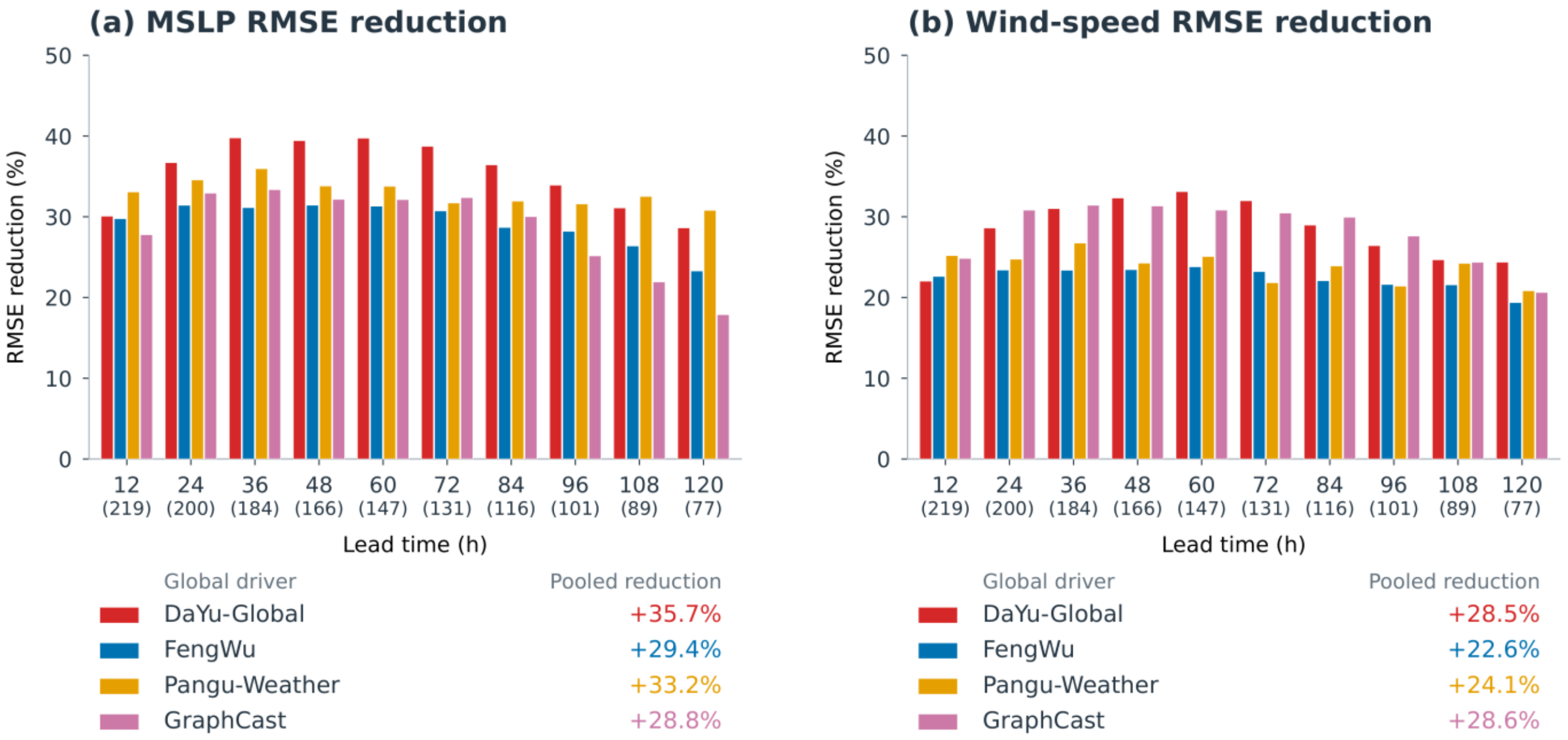
(a) MSLP RMSE reduction
RMSE reduction (%)
Lead time (h)
12 (219) 24 (200) 36 (184) 48 (166) 60 (147) 72 (131) 84 (116) 96 (101) 108 (89) 120 (77)
Global driver
Pooled reduction
DaYu-Global +35.7%
FengWu +29.4%
Pangu-Weather +33.2%
GraphCast +28.8%
(b) Wind-speed RMSE reduction
RMSE reduction (%)
Lead time (h)
Global driver
Pooled reduction
DaYu-Global +28.5%
FengWu +22.6%
Pangu-Weather +24.1%
GraphCast +28.6%

## Supplementary data

One Supplementary Data file accompanies this article. It contains Supplementary Results (Figs. S1–S13), Supplementary Methods (Fig. S14, Tables S1 and S2 and Eqs. S1–S12) and a linked contents list. Reference numbers correspond to those in the main article.

## Data availability

ERA5 reanalysis data are available from the ECMWF Climate Data Store (https://cds.climate.copernicus.eu/). IBTrACS Version 4r01 TC best-track data are available from NOAA's National Centers for Environmental Information (NCEI) (https://doi.org/10.25921/82ty-9e16). Historical IFS-HRES analyses and forecasts can be requested through the ECMWF Meteorological Archival and Retrieval System (MARS; https://www.ecmwf.int/en/forecasts/access-forecasts/access-archive-datasets).

## Code availability

The DaYu-Global and DaYu-Region architecture code, data-preprocessing code, inference entry points and instructions for global–regional coupling are publicly available at https://github.com/hourzp/DaYu-TC. The associated pretrained weights, inference configurations, normalization statistics, static inputs and usage instructions are archived on Zenodo (https://doi.org/10.5281/zenodo.23181947). Model weights are released under the CC BY-NC 4.0 licence. The accompanying release is v0.1.0-rc.1. Inference implementations, pretrained-weight access instructions and runtime requirements for the comparison models Pangu-Weather, FuXi, FengWu and GraphCast are available from their respective official code repositories.

# Supplementary Data

*DaYu-TC: a global–regional nested AI system for tropical cyclone forecasting*

Runzepu Hou[1,2], Feng Zhang[1,2*], Xin Wu[1,2], Xujun Wei[1,2], Fenghua Ling[3], Kan Dai[4], Gaozhen Nie[4,5], Leiming Ma[6], Renhe Zhang[1]

[1] Key Laboratory of Polar Atmosphere-Ocean-Ice System for Weather and Climate of Ministry of Education / Shanghai Key Laboratory of Ocean-Land-Atmosphere Boundary Dynamics and Climate Change, Department of Atmospheric and Oceanic Sciences & Institutes of Atmospheric Sciences, Fudan University, Shanghai 200438, China

[2] Key Laboratory for Information Science of Electromagnetic Waves, Ministry of Education, College of Future Information Technology, Fudan University, Shanghai 200438, China

[3] Shanghai Artificial Intelligence Laboratory, Shanghai 200232, China

[4] National Meteorological Center, Beijing 100081, China

[5] Xiong'an Institute of Meteorological Artificial Intelligence, Hebei 070001, China

[6] Shanghai Typhoon Institute, China Meteorological Administration, Shanghai 200030, China

* Correspondence: Feng Zhang; Email: fengzhang@fudan.edu.cn

This supplement contains additional forecast evaluations and case studies, followed by detailed data, model, training and evaluation methods. Figures, tables and equations are numbered separately. Reference numbers follow those in the main article.

# Contents

## Supplementary Results

### Conventional global forecast skill

The global evaluations complement the extreme-anomaly and tropical cyclone assessments in the main text. Across the eight variables shown in Supplementary Fig. S1, DaYu-Global maintains forecast skill comparable to FuXi and FengWu, with lower latitude-weighted root-mean-square error (wRMSE) and higher latitude-weighted anomaly correlation coefficient (wACC) than GraphCast and Pangu-Weather for several variables at medium-to-long lead times. Thus, the improved representation of strong-wind and low-pressure anomalies is accompanied by competitive conventional weather forecast skill. The forecast initialized at 00:00 UTC on 1 August 2023 further illustrates the retention of major circulation patterns and broad spatial structures through day 10 across surface and upper-air fields (Supplementary Fig. S2).

### Storm-specific global intensity forecasts

The storm-specific comparison shows how maximum-wind errors vary across the 16 evaluated tropical cyclones (TCs) in the western North Pacific (WNP) (Supplementary Fig. S3). DaYu-Global achieves the lowest root-mean-square error (RMSE) for multiple storms spanning the displayed peak-intensity groups, including Sanba, Talim, Lan, Khanun, Mawar and Bolaven. Cell colours encode within-storm rankings, whereas the numerical values retain the absolute error magnitudes. The peak-wind bars provide intensity context rather than an additional forecast metric.

### Global tropical cyclone case studies

Global wind-field and track comparisons are shown for Koinu, Doksuri and Bolaven (Supplementary Figs. S4–S6). In the Koinu case, DaYu-Global has the smallest absolute maximum-wind and minimum-pressure biases among the evaluated global AI models on days 2–5. Its day-5 biases are −15.9 m $s^{-1}$ and +18.2 hPa, respectively. The Doksuri and Bolaven cases provide further comparisons of near-centre wind structures and their evolution. All forecast biases are evaluated against the same NOAA IBTrACS best-track reference.

### Regional tropical cyclone case studies

The storm-centred Doksuri and Bolaven comparisons illustrate how regional nesting modifies near-centre wind and pressure structures (Supplementary Figs. S7 and S8). On day 4 of Doksuri, the DaYu-TC wind-speed and pressure biases are −16.1 m $s^{-1}$ and +8.2 hPa, compared with −28.0 m $s^{-1}$ and +26.5 hPa for DaYu-Global. In Bolaven, DaYu-TC produces a more concentrated near-centre strong-wind region on days 2–3. These additional cases connect the intensity gains to their spatial expression, while the separately reported centre displacement distinguishes position errors from the storm-centred structural comparisons.

## Conventional regional forecast skill

Regional nesting also improves conventional atmospheric fields over the verification domain (Supplementary Fig. S9). At 6 h, RMSE decreases for all nine variables, with reductions of 17.5–44.6%; the largest reduction occurs in mean sea-level pressure. Positive RMSE and ACC improvements persist throughout 120 h for T2M, MSLP and T850. These results show that the tropical cyclone intensity gains are accompanied by broader improvements in short-range regional forecasts, with the persistence of the gains depending on the variable and lead time.

## Wind-field refinement across global drivers

Paired forecasts for Doksuri, Koinu and Bolaven connect the aggregate cross-driver intensity gains in the main text to their spatial expression (Supplementary Figs. S10–S12). Under Pangu-Weather, GraphCast and FengWu backgrounds, regional forecasts generally produce stronger and more concentrated near-centre wind fields, with smaller negative maximum-wind biases than their global drivers. For Bolaven on day 3, the wind-speed biases decrease in magnitude from 54.4, 56.4 and 56.2 m $s^{-1}$ to 38.7, 38.6 and 41.2 m $s^{-1}$, respectively. The same fixed regional weights are used throughout, linking these structural refinements to the cross-driver gains without driver-specific retraining.

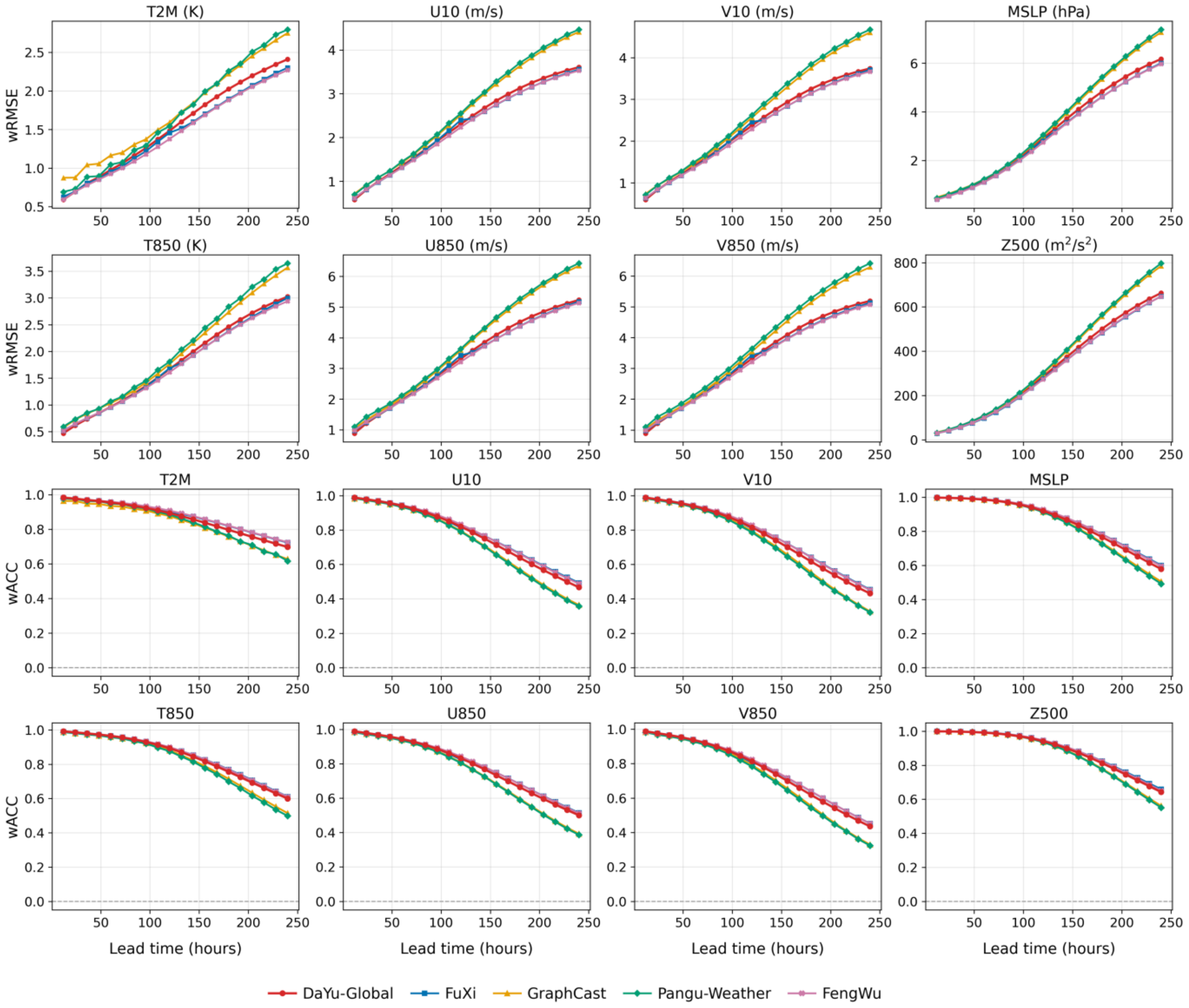

Supplementary Figure S1. Global deterministic forecast performance on the independent 2023 test set. Latitude-weighted root-mean-square error (wRMSE; upper two rows) and latitude-weighted anomaly correlation coefficient (wACC; lower two rows) are evaluated against ERA5 from 12 to 240 h. The eight variables are 2 m temperature (T2M), 10 m zonal and meridional wind (U10 and V10), mean sea-level pressure (MSLP), 850 hPa temperature and zonal and meridional wind (T850, U850 and V850) and 500 hPa geopotential (Z500). Curves compare DaYu-Global with FuXi, GraphCast, Pangu-Weather and FengWu. Lower wRMSE and higher wACC indicate better forecast skill; the units of wRMSE are given in the corresponding panel titles.

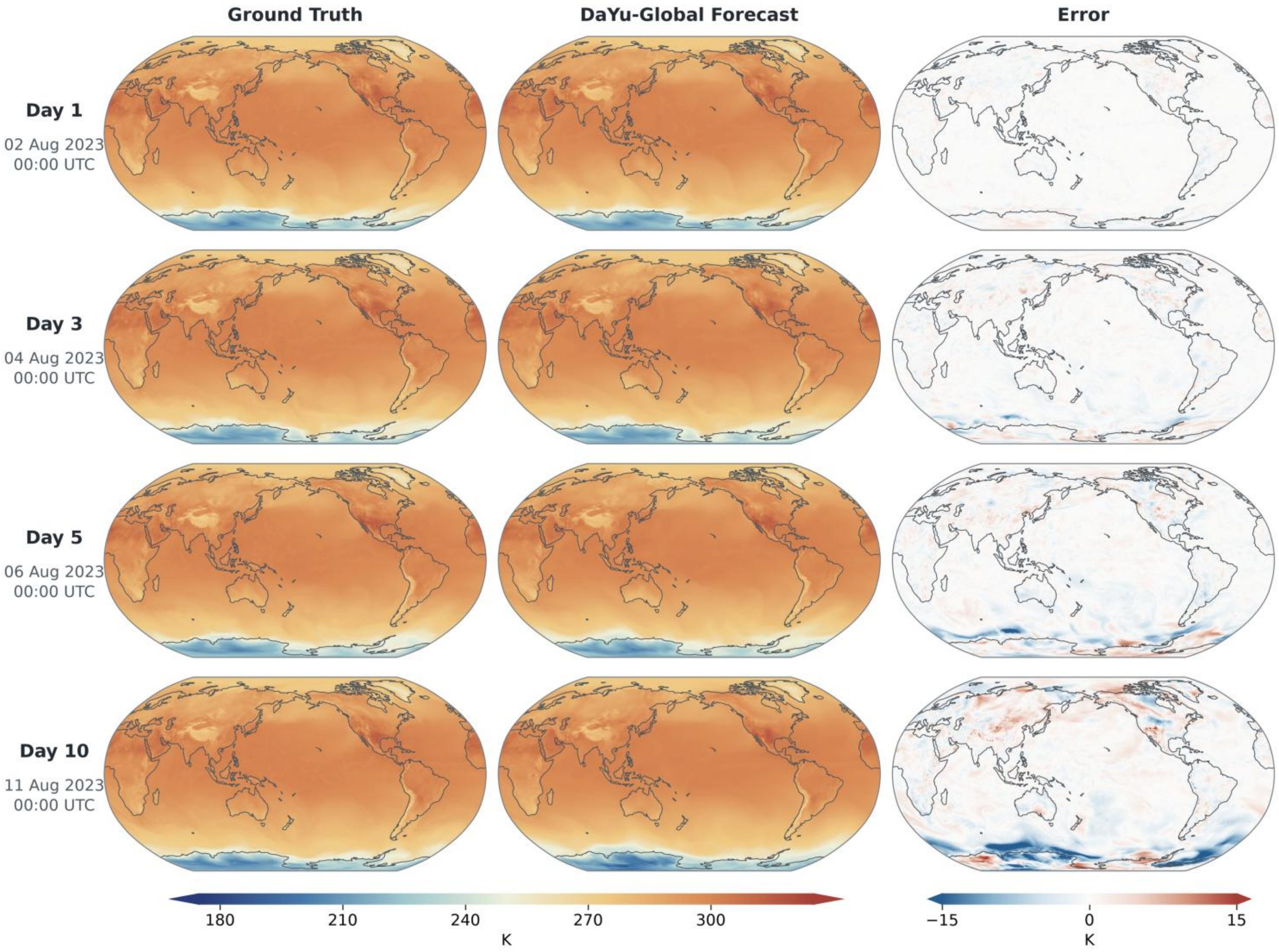

Supplementary Figure S2a. Spatial evolution of DaYu-Global forecasts: 2 m temperature. Forecasts are initialized at 00:00 UTC on 1 August 2023. Rows show days 1, 3, 5 and 10 (24, 72, 120 and 240 h), respectively. Columns show the verifying ERA5 field (labelled 'Ground Truth'), the DaYu-Global forecast and the corresponding forecast error. Colour bars give T2M and its error in K. Panels S2a–h together constitute Supplementary Fig. S2.

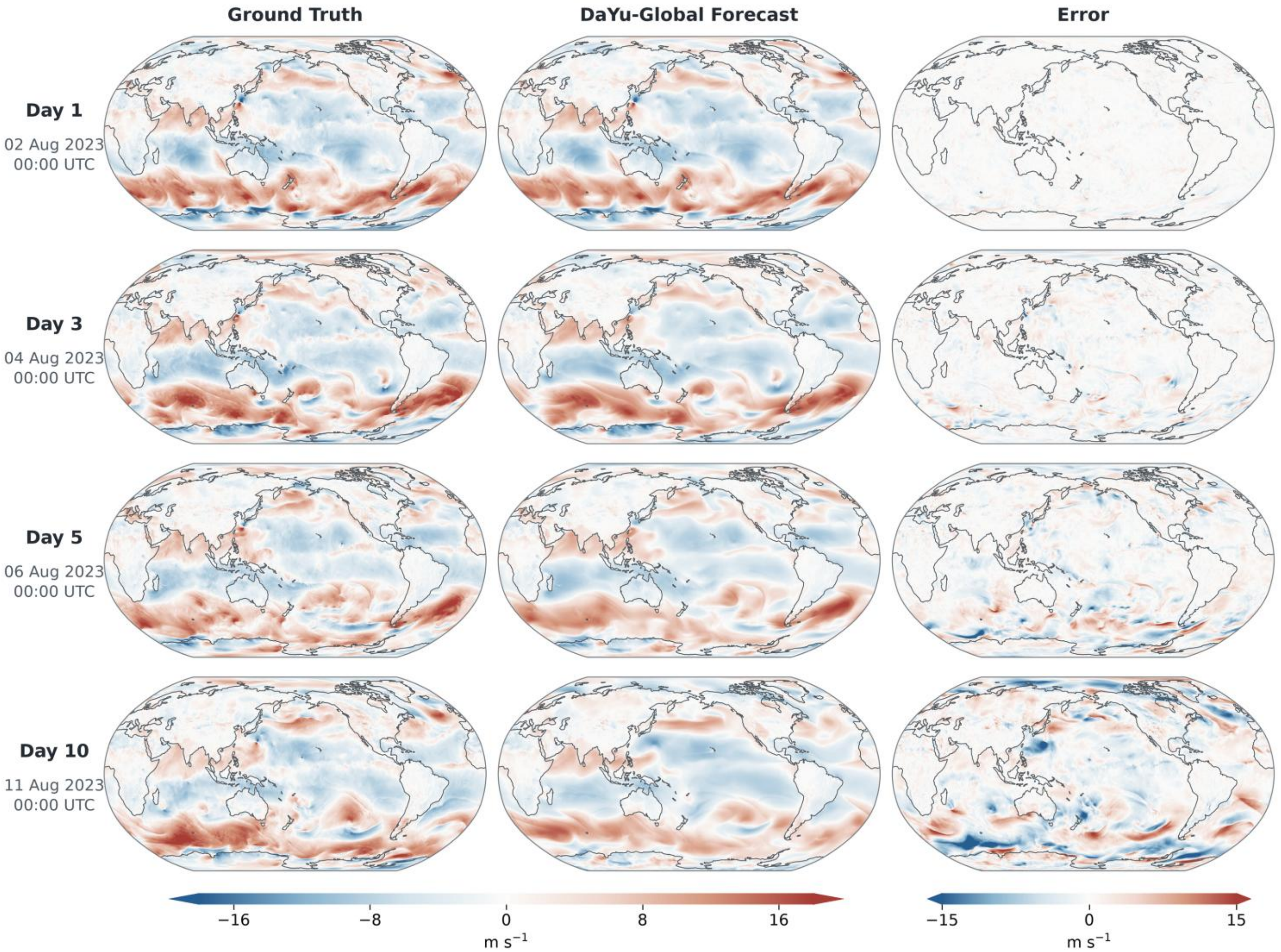

Supplementary Figure S2b. Spatial evolution of DaYu-Global forecasts: 10 m zonal wind. Forecasts are initialized at 00:00 UTC on 1 August 2023. Rows show days 1, 3, 5 and 10 (24, 72, 120 and 240 h), respectively. Columns show the verifying ERA5 field (labelled ‘Ground Truth’), the DaYu-Global forecast and the corresponding forecast error. Colour bars give U10 and its error in m $s^{-1}$. Panels S2a–h together constitute Supplementary Fig. S2.

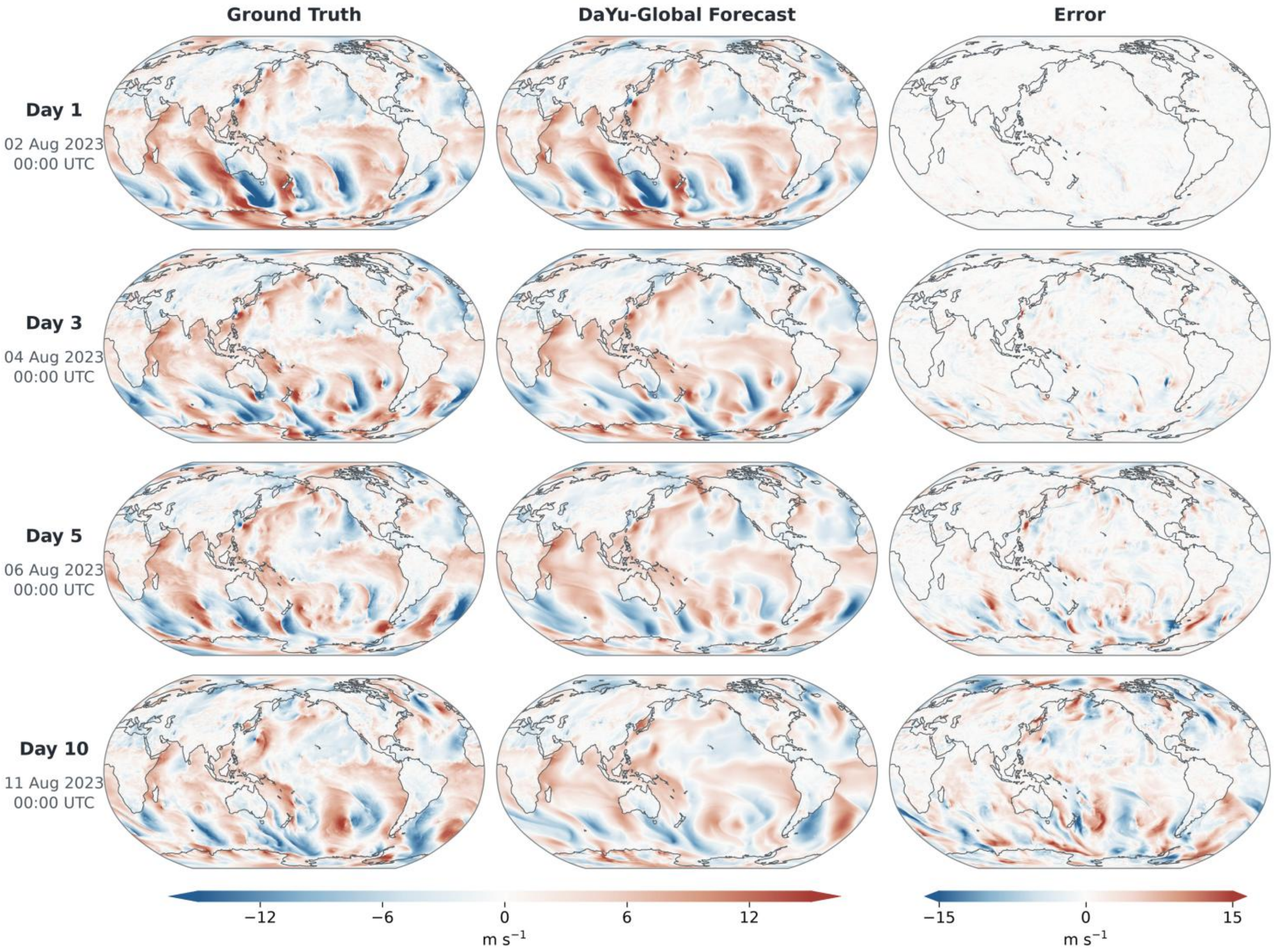

Supplementary Figure S2c. Spatial evolution of DaYu-Global forecasts: 10 m meridional wind. Forecasts are initialized at 00:00 UTC on 1 August 2023. Rows show days 1, 3, 5 and 10 (24, 72, 120 and 240 h), respectively. Columns show the verifying ERA5 field (labelled ‘Ground Truth’), the DaYu-Global forecast and the corresponding forecast error. Colour bars give V10 and its error in m $s^{-1}$. Panels S2a–h together constitute Supplementary Fig. S2.

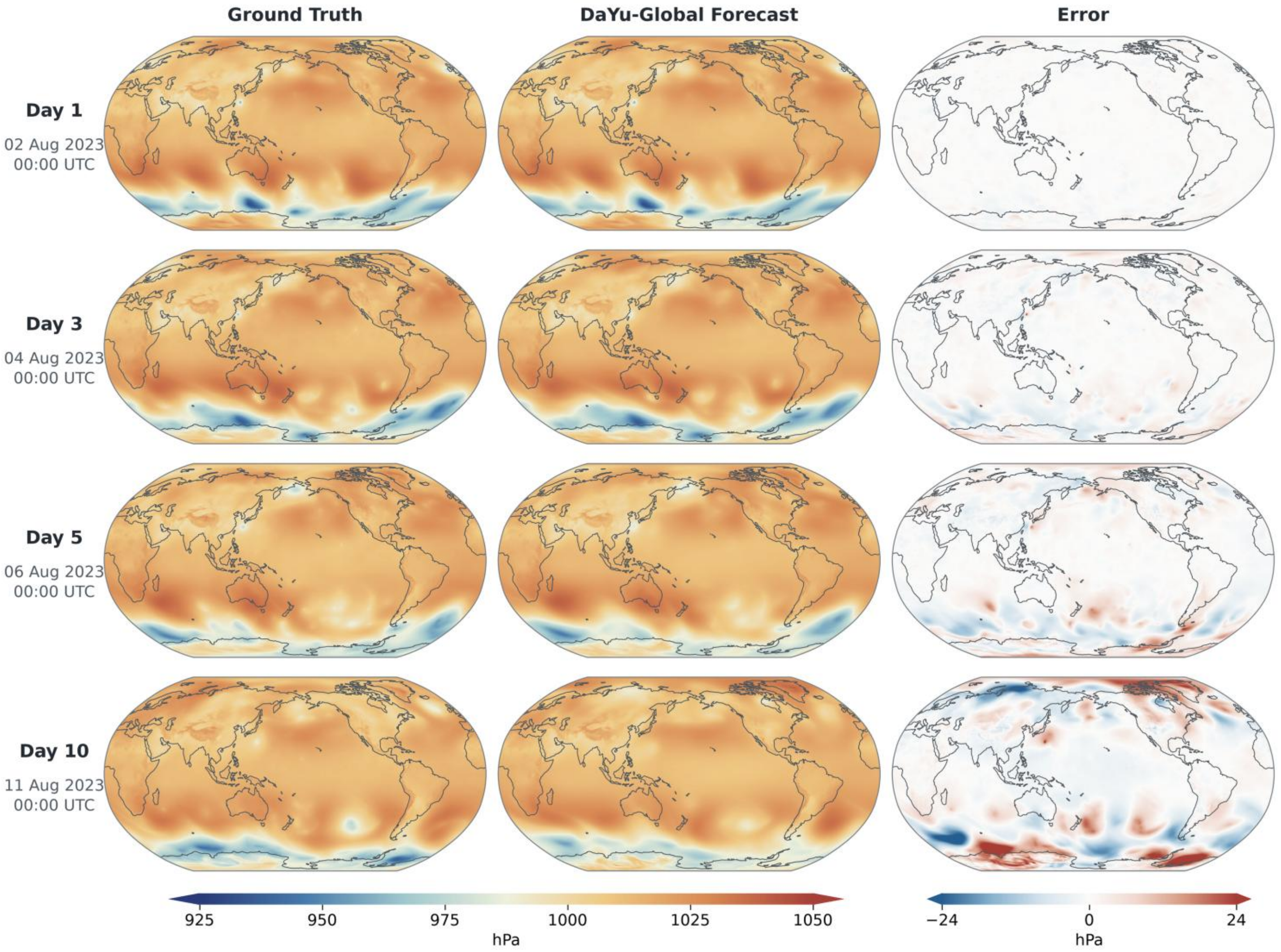

Supplementary Figure S2d. Spatial evolution of DaYu-Global forecasts: mean sea-level pressure. Forecasts are initialized at 00:00 UTC on 1 August 2023. Rows show days 1, 3, 5 and 10 (24, 72, 120 and 240 h), respectively. Columns show the verifying ERA5 field (labelled 'Ground Truth'), the DaYu-Global forecast and the corresponding forecast error. Colour bars give MSLP and its error in hPa. Panels S2a–h together constitute Supplementary Fig. S2.

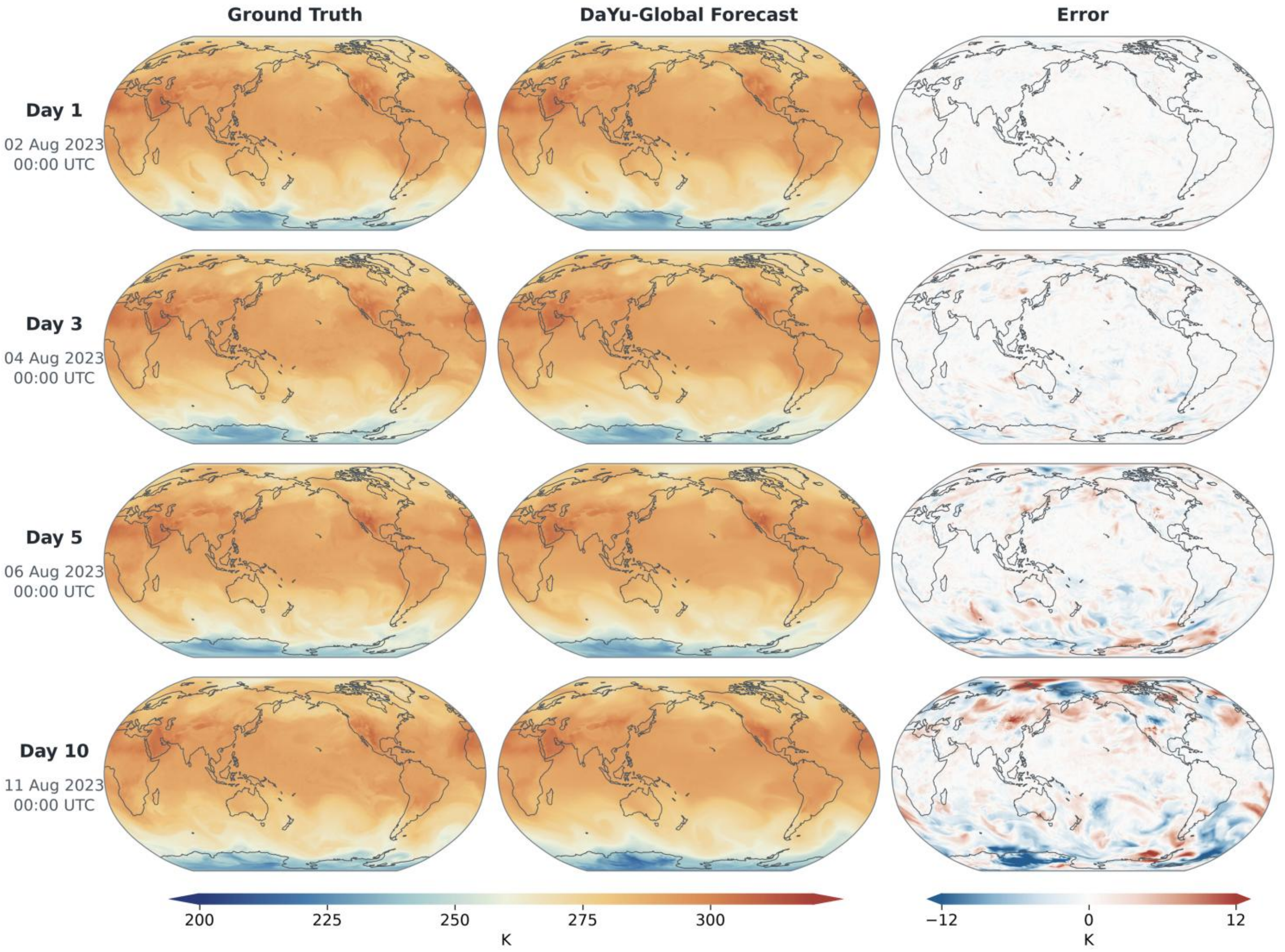

Supplementary Figure S2e. Spatial evolution of DaYu-Global forecasts: 850 hPa temperature. Forecasts are initialized at 00:00 UTC on 1 August 2023. Rows show days 1, 3, 5 and 10 (24, 72, 120 and 240 h), respectively. Columns show the verifying ERA5 field (labelled 'Ground Truth'), the DaYu-Global forecast and the corresponding forecast error. Colour bars give T850 and its error in K. Panels S2a–h together constitute Supplementary Fig. S2.

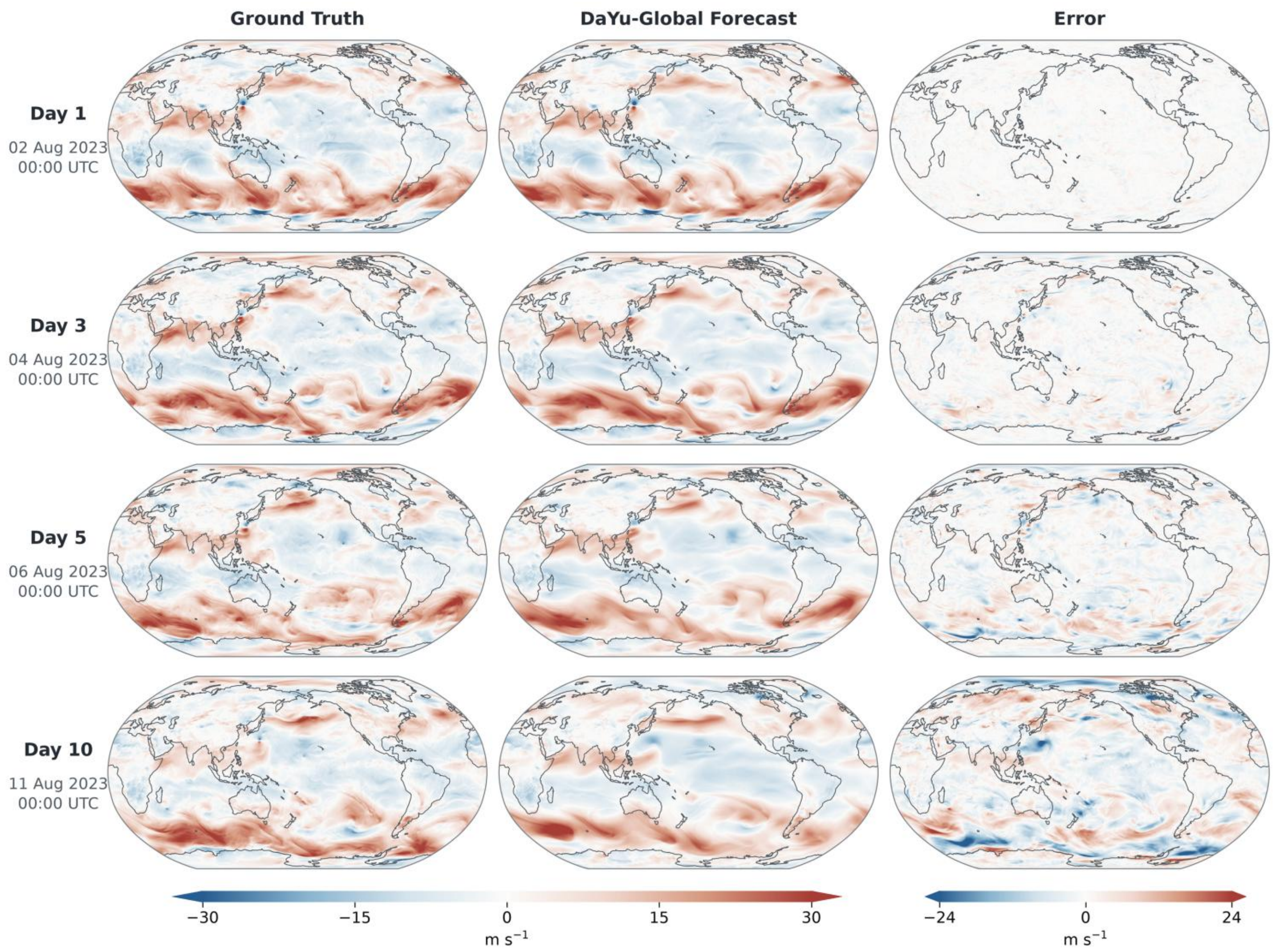

Supplementary Figure S2f. Spatial evolution of DaYu-Global forecasts: 850 hPa zonal wind. Forecasts are initialized at 00:00 UTC on 1 August 2023. Rows show days 1, 3, 5 and 10 (24, 72, 120 and 240 h), respectively. Columns show the verifying ERA5 field (labelled ‘Ground Truth’), the DaYu-Global forecast and the corresponding forecast error. Colour bars give U850 and its error in m $s^{-1}$. Panels S2a–h together constitute Supplementary Fig. S2.

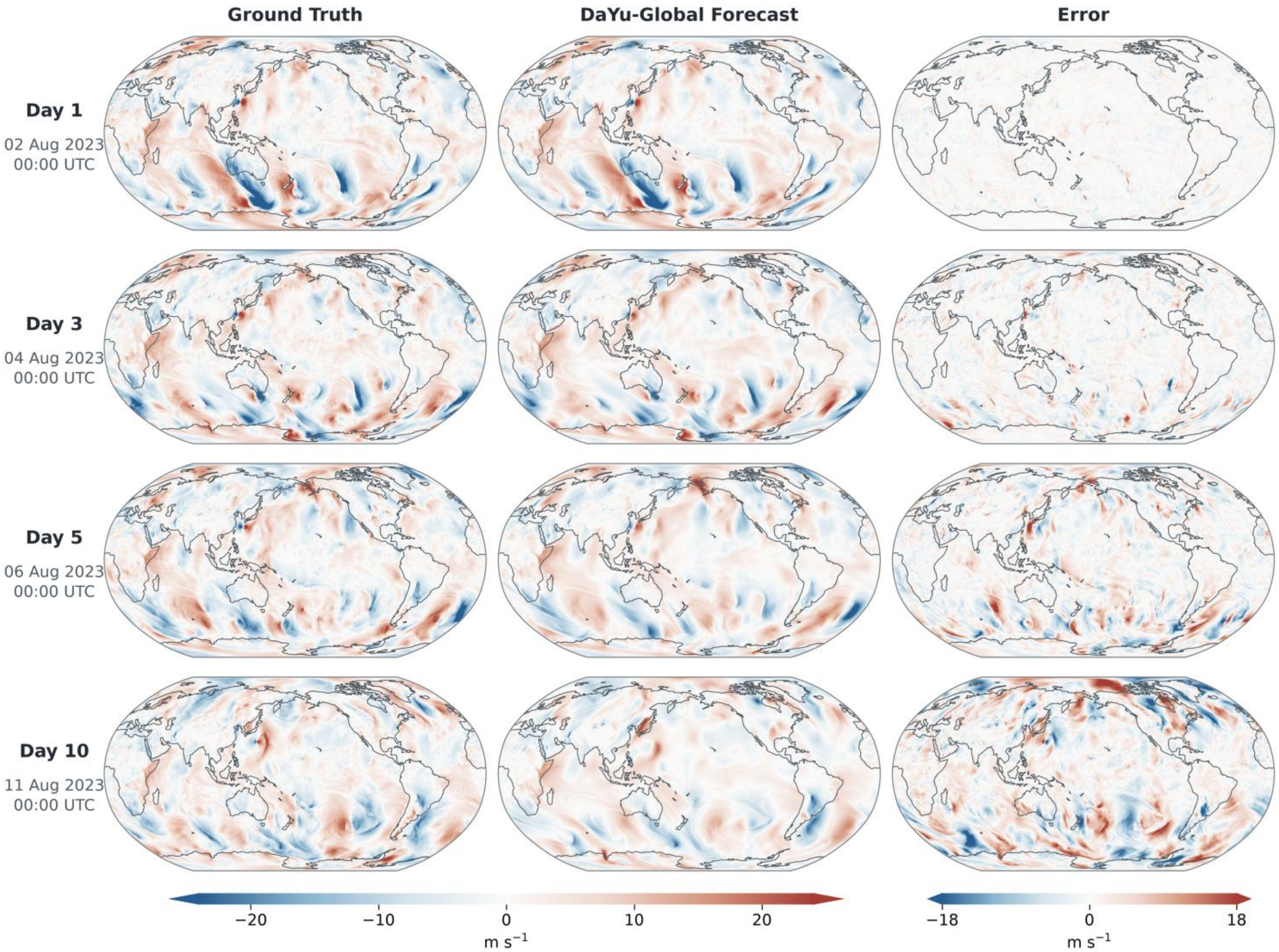

Supplementary Figure S2g. Spatial evolution of DaYu-Global forecasts: 850 hPa meridional wind. Forecasts are initialized at 00:00 UTC on 1 August 2023. Rows show days 1, 3, 5 and 10 (24, 72, 120 and 240 h), respectively. Columns show the verifying ERA5 field (labelled 'Ground Truth'), the DaYu-Global forecast and the corresponding forecast error. Colour bars give V850 and its error in m $s^{-1}$. Panels S2a–h together constitute Supplementary Fig. S2.

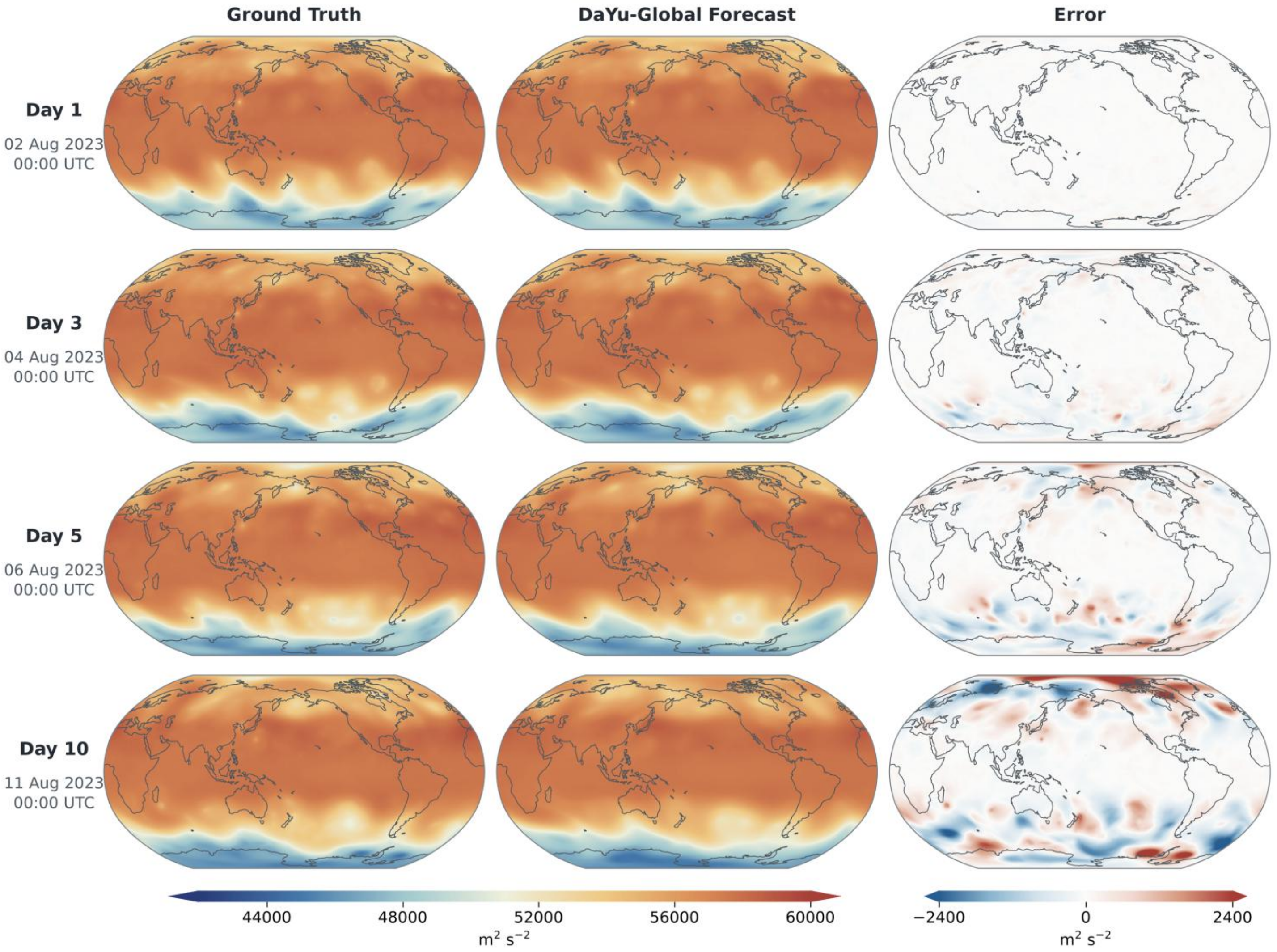

Supplementary Figure S2h. Spatial evolution of DaYu-Global forecasts: 500 hPa geopotential. Forecasts are initialized at 00:00 UTC on 1 August 2023. Rows show days 1, 3, 5 and 10 (24, 72, 120 and 240 h), respectively. Columns show the verifying ERA5 field (labelled 'Ground Truth'), the DaYu-Global forecast and the corresponding forecast error. Colour bars give Z500 and its error in $m^2 s^{-2}$. Panels S2a–h together constitute Supplementary Fig. S2.

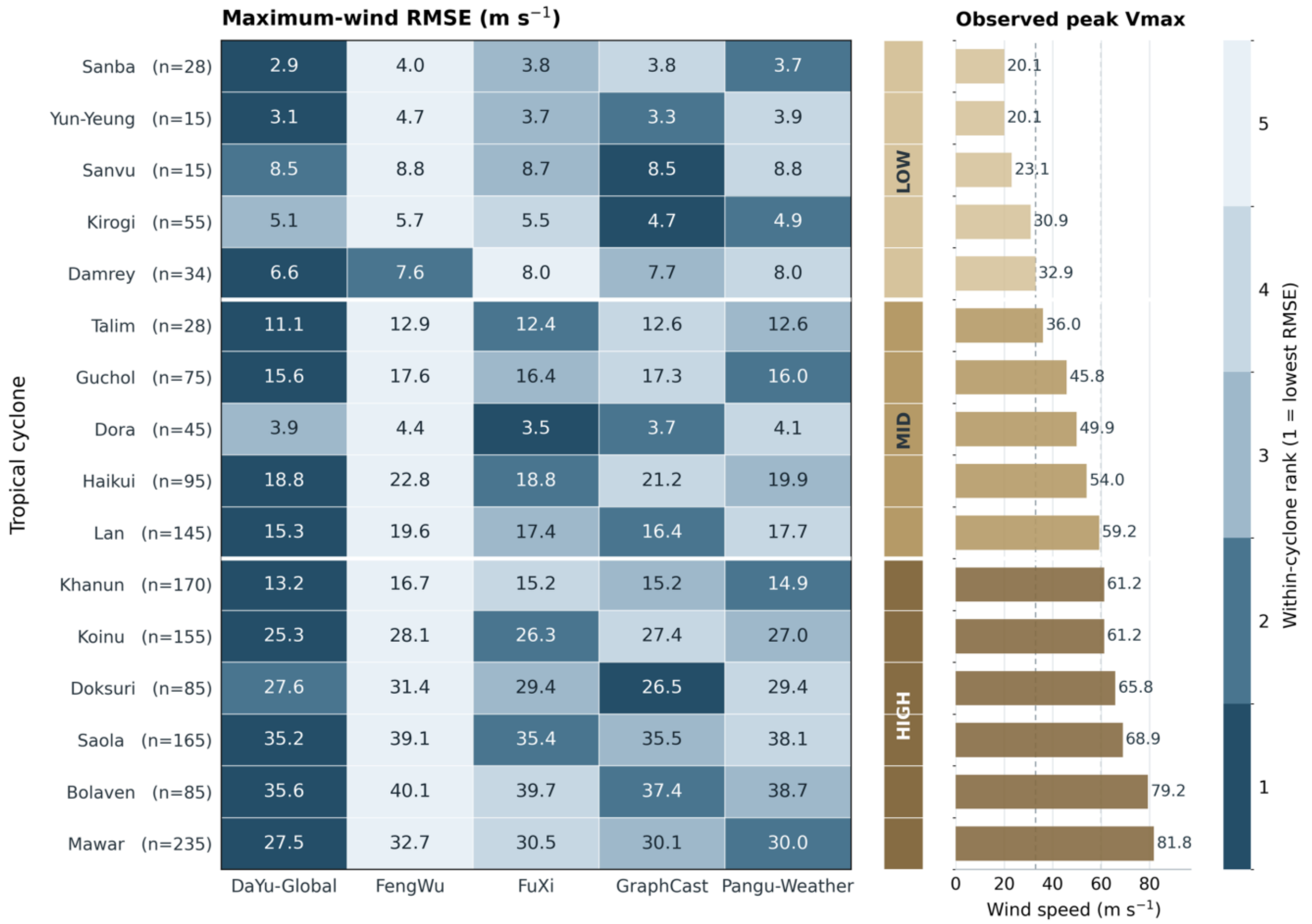

| Tropical cyclone | DaYu-Global | FengWu | FuXi | GraphCast | Pangu-Weather | Observed peak Vmax |
|---|---|---|---|---|---|---|
| Sanba (n=28) | 2.9 | 4.0 | 3.8 | 3.8 | 3.7 | 20.1 |
| Yun-Yeung (n=15) | 3.1 | 4.7 | 3.7 | 3.3 | 3.9 | 20.1 |
| Sanvu (n=15) | 8.5 | 8.8 | 8.7 | 8.5 | 8.8 | 23.1 |
| Kirogi (n=55) | 5.1 | 5.7 | 5.5 | 4.7 | 4.9 | 30.9 |
| Damrey (n=34) | 6.6 | 7.6 | 8.0 | 7.7 | 8.0 | 32.9 |
| Talim (n=28) | 11.1 | 12.9 | 12.4 | 12.6 | 12.6 | 36.0 |
| Guchol (n=75) | 15.6 | 17.6 | 16.4 | 17.3 | 16.0 | 45.8 |
| Dora (n=45) | 3.9 | 4.4 | 3.5 | 3.7 | 4.1 | 49.9 |
| Haikui (n=95) | 18.8 | 22.8 | 18.8 | 21.2 | 19.9 | 54.0 |
| Lan (n=145) | 15.3 | 19.6 | 17.4 | 16.4 | 17.7 | 59.2 |
| Khanun (n=170) | 13.2 | 16.7 | 15.2 | 15.2 | 14.9 | 61.2 |
| Koinu (n=155) | 25.3 | 28.1 | 26.3 | 27.4 | 27.0 | 61.2 |
| Doksuri (n=85) | 27.6 | 31.4 | 29.4 | 26.5 | 29.4 | 65.8 |
| Saola (n=165) | 35.2 | 39.1 | 35.4 | 35.5 | 38.1 | 68.9 |
| Bolaven (n=85) | 35.6 | 40.1 | 39.7 | 37.4 | 38.7 | 79.2 |
| Mawar (n=235) | 27.5 | 32.7 | 30.5 | 30.1 | 30.0 | 81.8 |

Supplementary Figure S3. Storm-specific maximum-wind forecast performance of global AI models. Cells show maximum 10 m wind-speed RMSE (m $s^{-1}$) for 16 TCs in the WNP in 2023, evaluated against NOAA IBTrACS best-track estimates. Columns compare DaYu-Global, FengWu, FuXi, GraphCast and Pangu-Weather. Numbers in parentheses after storm names indicate valid sample counts. Cell colours encode the model rank within each storm, with rank 1 indicating the lowest RMSE; the colour scale does not represent absolute RMSE. Rankings are based on unrounded RMSE values. Storms are ordered by the peak best-track wind speeds shown in the right-hand bars and grouped into the low-, medium- and high-intensity bands displayed in the figure.

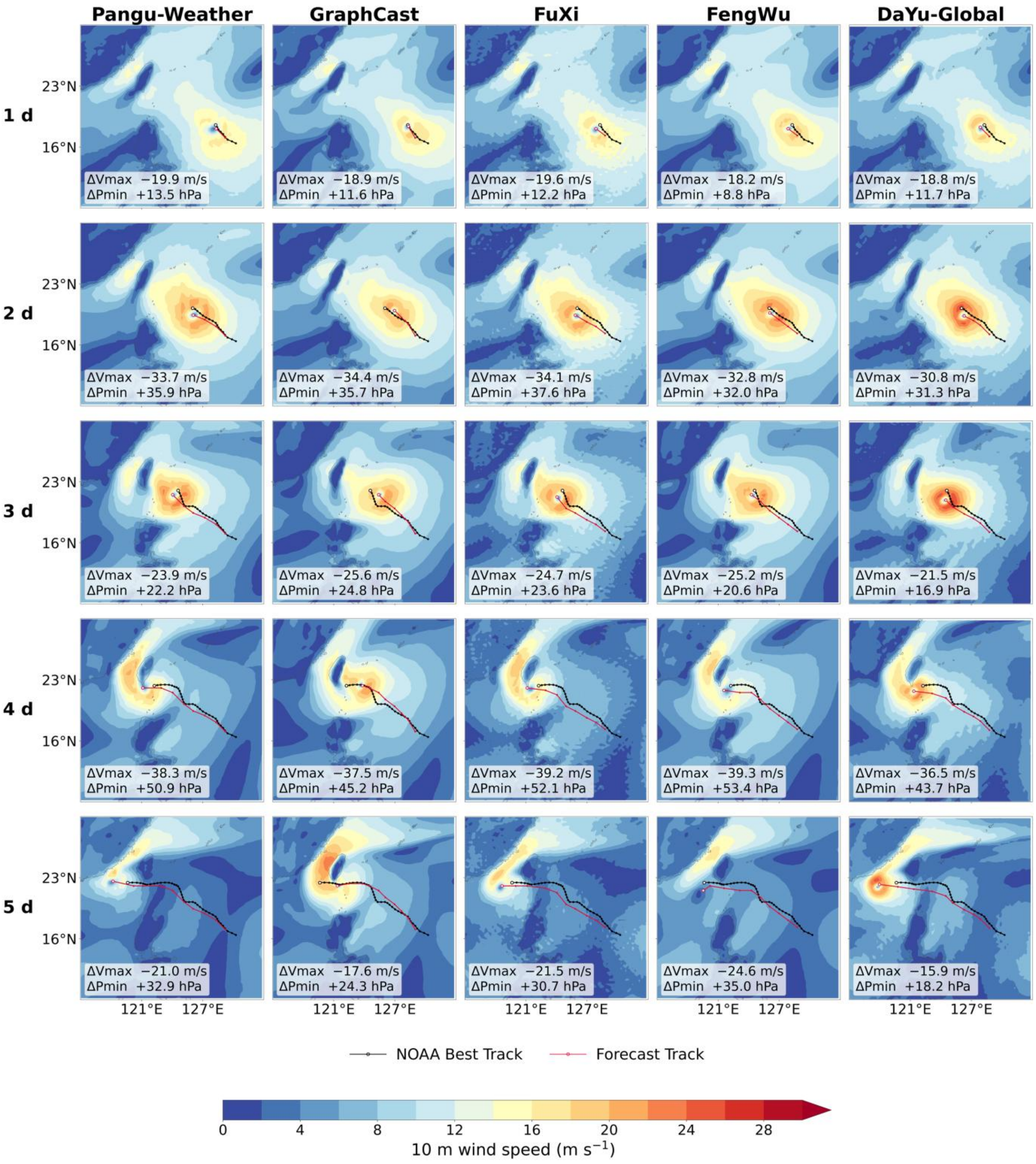

Supplementary Figure S4. Wind-field structures and track forecasts for TC Koinu from global AI models. Forecasts initialized at 12:00 UTC on 30 September 2023 are shown at lead times of 1–5 d. Columns show Pangu-Weather, GraphCast, FuXi, FengWu and DaYu-Global; rows correspond to days 1–5. Shading indicates 10 m wind speed (m $s^{-1}$). Black lines show the NOAA IBTrACS best track and red lines show forecast tracks. Open circles mark centres at the corresponding valid time. ΔVmax and ΔPmin denote maximum 10 m wind-speed and minimum mean sea-level pressure biases, in m $s^{-1}$ and hPa, respectively, calculated as the forecast minus the NOAA best-track estimate. Negative wind-speed biases and positive pressure biases indicate underestimated intensity.

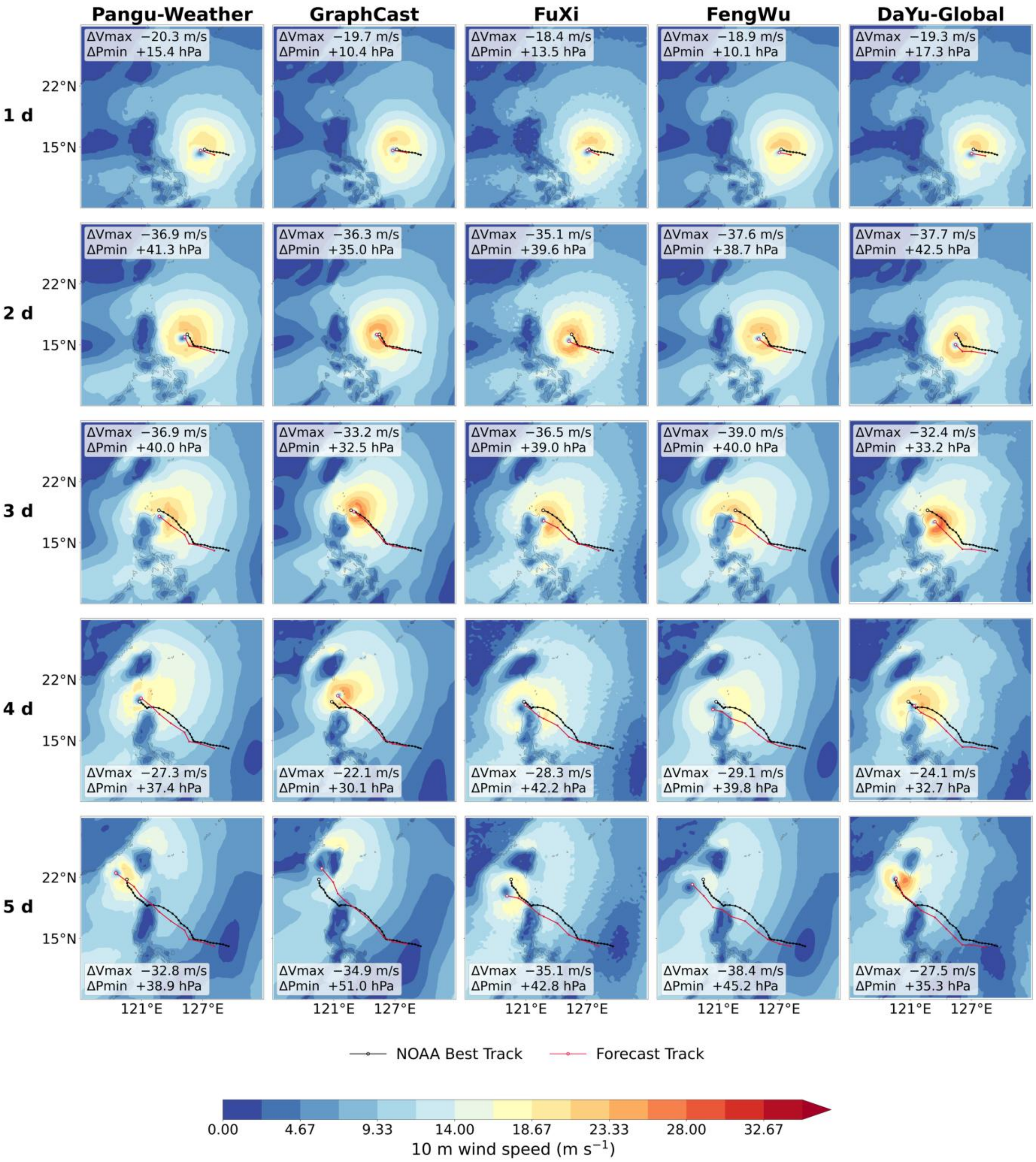

Supplementary Figure S5. Global AI forecasts of tropical cyclone Doksuri. Forecasts initialized at 12:00 UTC on 22 July 2023 are shown at 1–5 d (rows). Columns show Pangu-Weather, GraphCast, FuXi, FengWu and DaYu-Global. Shading indicates 10 m wind speed (m s$^{-1}$). Black and red curves show NOAA best tracks and forecast tracks, respectively; open circles mark the centres at the valid time. Insets report ΔVmax (m s$^{-1}$) and ΔPmin (hPa), defined as forecast maximum 10 m wind speed and minimum mean sea-level pressure minus the corresponding NOAA best-track estimates. Negative ΔVmax and positive ΔPmin indicate underestimated tropical cyclone intensity.

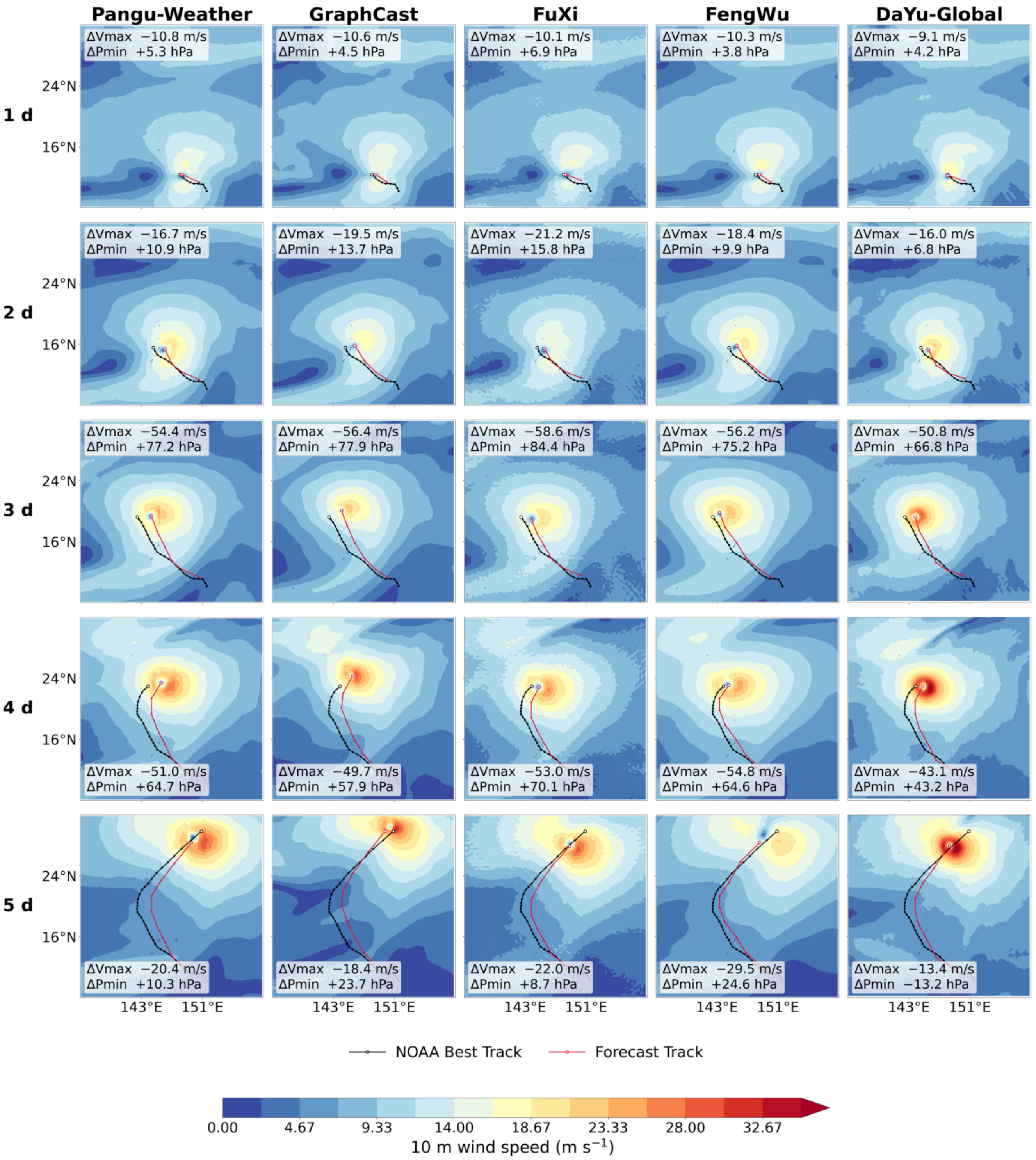

Supplementary Figure S6. Global AI forecasts of tropical cyclone Bolaven. Forecasts initialized at 12:00 UTC on 8 October 2023 are shown at 1–5 d (rows). Columns show Pangu-Weather, GraphCast, FuXi, FengWu and DaYu-Global. Shading indicates 10 m wind speed (m $s^{-1}$). Black and red curves show NOAA best tracks and forecast tracks, respectively; open circles mark the centres at the valid time. Insets report ΔVmax (m $s^{-1}$) and ΔPmin (hPa), defined as forecast maximum 10 m wind speed and minimum mean sea-level pressure minus the corresponding NOAA best-track estimates. Negative ΔVmax and positive ΔPmin indicate underestimated tropical cyclone intensity.

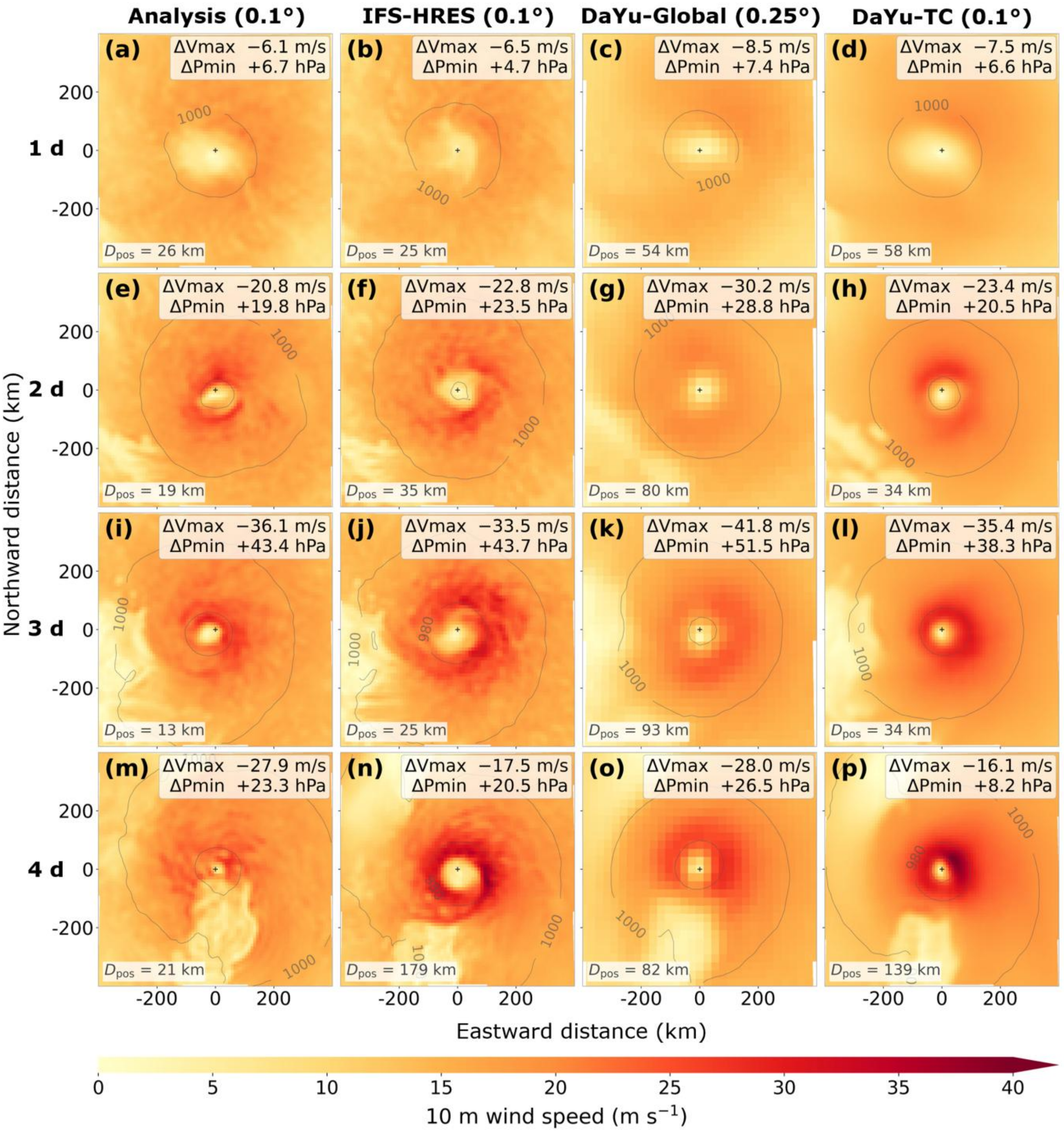

Supplementary Figure S7. Regional refinement of tropical cyclone Doksuri by DaYu-TC. Forecasts initialized at 00:00 UTC on 22 July 2023 are shown at 1–4 d (rows; panels a–p), together with the IFS analysis at the corresponding valid times. Columns show the IFS analysis (Analysis), IFS-HRES, DaYu-Global and DaYu-TC. Shading indicates 10 m wind speed (m s$^{-1}$) and grey contours show mean sea-level pressure (hPa). Each panel is centred on its own diagnosed tropical cyclone centre, marked by a cross; the horizontal and vertical axes give eastward and northward distances (km). Insets report ΔVmax (m s$^{-1}$) and ΔPmin (hPa), defined as the diagnosed or forecast maximum 10 m wind speed and minimum mean sea-level pressure minus the corresponding NOAA best-track estimates. Dpos is the distance between the diagnosed centre and the NOAA best-track centre at the same valid time (km).

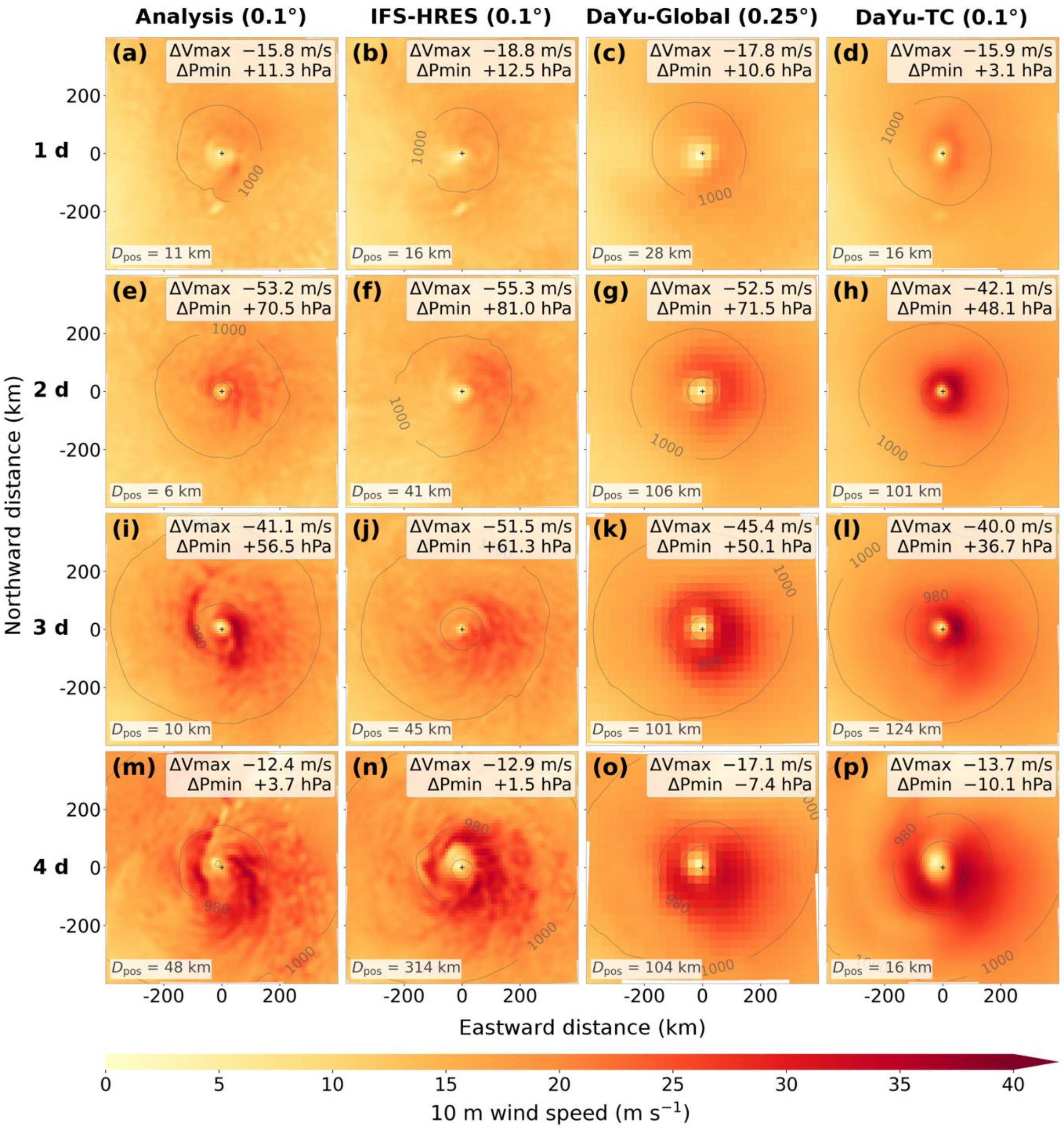

Supplementary Figure S8. Regional refinement of tropical cyclone Bolaven by DaYu-TC. Forecasts initialized at 12:00 UTC on 9 October 2023 are shown at 1–4 d (rows; panels a–p), together with the IFS analysis at the corresponding valid times. Columns show the IFS analysis (Analysis), IFS-HRES, DaYu-Global and DaYu-TC. Shading indicates 10 m wind speed (m $s^{-1}$) and grey contours show mean sea-level pressure (hPa). Each panel is centred on its own diagnosed tropical cyclone centre, marked by a cross; the horizontal and vertical axes give eastward and northward distances (km). Insets report ΔVmax (m $s^{-1}$) and ΔPmin (hPa), defined as the diagnosed or forecast maximum 10 m wind speed and minimum mean sea-level pressure minus the corresponding NOAA best-track estimates. Dpos is the distance between the diagnosed centre and the NOAA best-track centre at the same valid time (km).

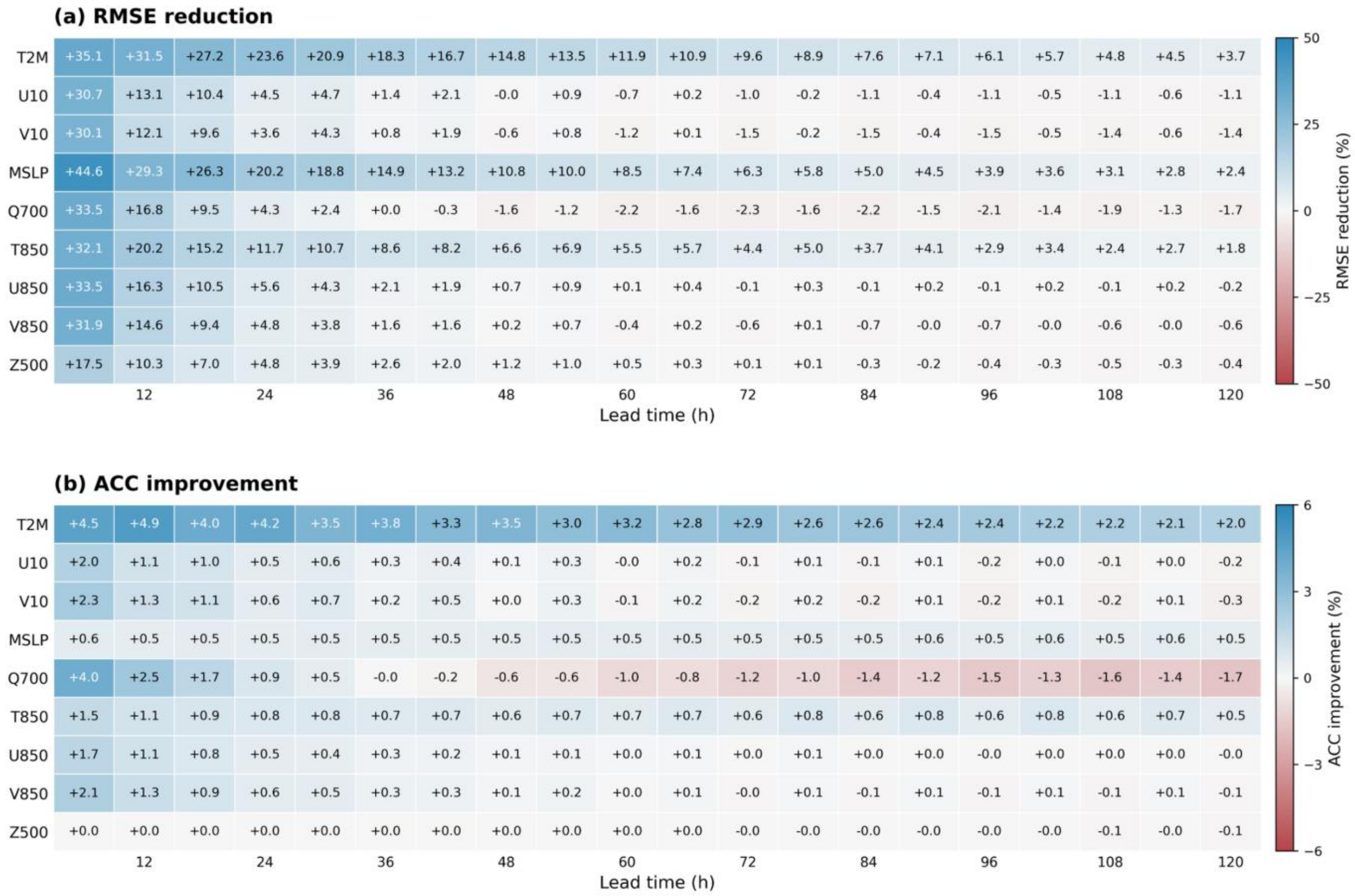

Supplementary Figure S9. Changes in conventional regional forecast skill after nesting DaYu-Global with DaYu-Region. Heatmaps show a, relative RMSE reductions and b, relative ACC increases for the default DaYu-TC configuration compared with DaYu-Global from 6 to 120 h. Values are calculated as 100 × (RMSEG − RMSER)/RMSEG and 100 × (ACCR − ACCG)/ACCG, respectively, where G and R denote global and regional forecasts. Positive values indicate improved skill and negative values indicate reduced skill. Rows show T2M, U10, V10, MSLP, 700 hPa specific humidity (Q700), T850, U850, V850 and Z500. Both forecasts are evaluated against the IFS analysis over the same verification domain and samples. Cell annotations are percentages rounded to one decimal place; values displayed as +0.0 or −0.0 reflect rounding.

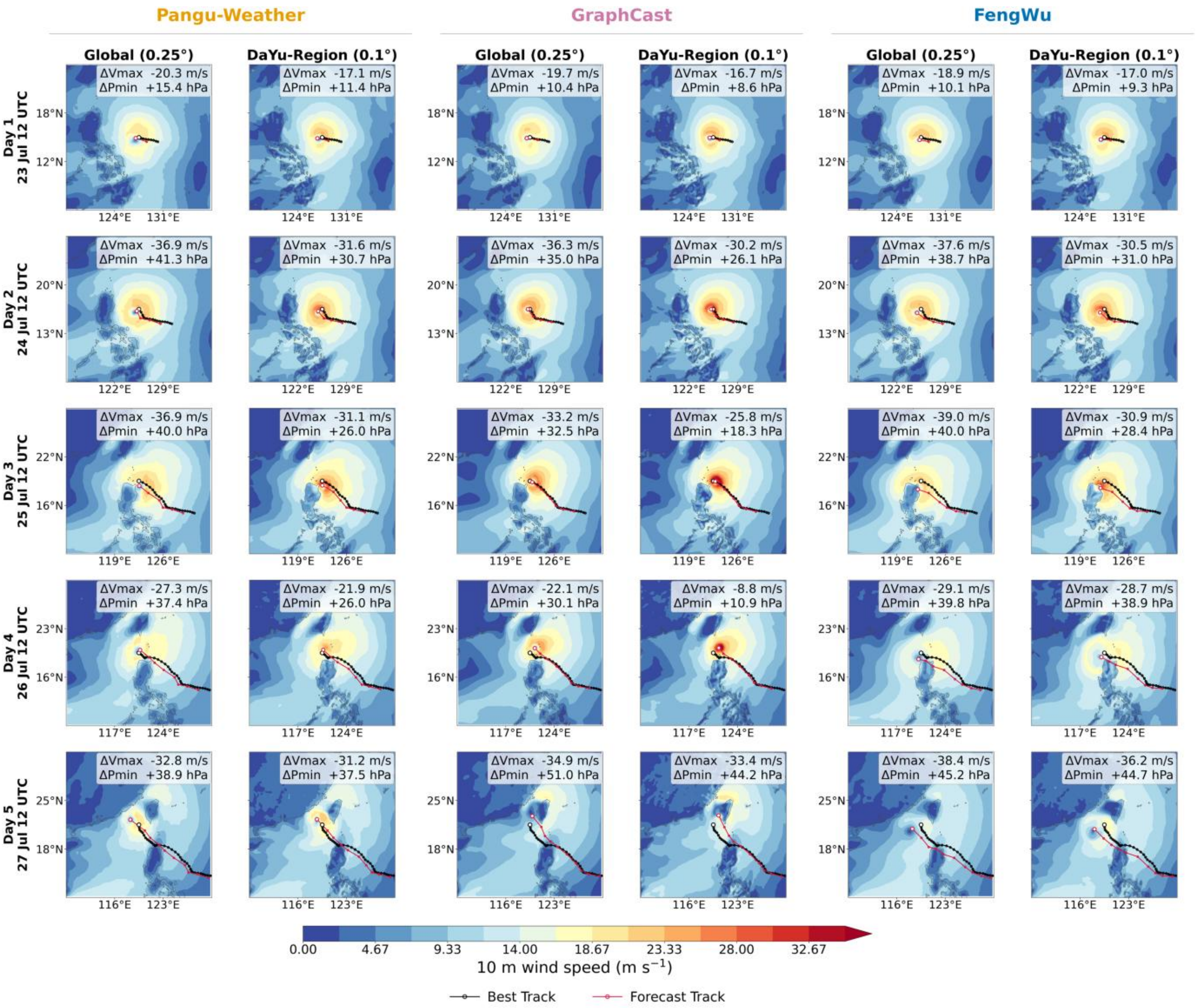

Supplementary Figure S10. Regional refinement of tropical cyclone Doksuri under different global forecast drivers. Forecasts initialized at 12:00 UTC on 22 July 2023 are shown at lead times of 1–5 d (rows). Column pairs correspond to Pangu-Weather, GraphCast and FengWu, from left to right. Within each pair, the global forecast (0.25°) is shown on the left and the corresponding DaYu-Region forecast (0.1°) on the right. All regional forecasts use the same fixed weights trained with DaYu-Global backgrounds, without driver-specific fine-tuning. Shading indicates 10 m wind speed (m $s^{-1}$). Black and red lines denote the NOAA IBTrACS best track and forecast track, respectively; open circles mark the centres at the corresponding valid time. Annotations give ΔVmax and ΔPmin, defined as forecast minus best-track estimates of maximum 10 m wind speed (m $s^{-1}$) and minimum mean sea-level pressure (hPa), respectively. Negative ΔVmax and positive ΔPmin indicate underestimated intensity.

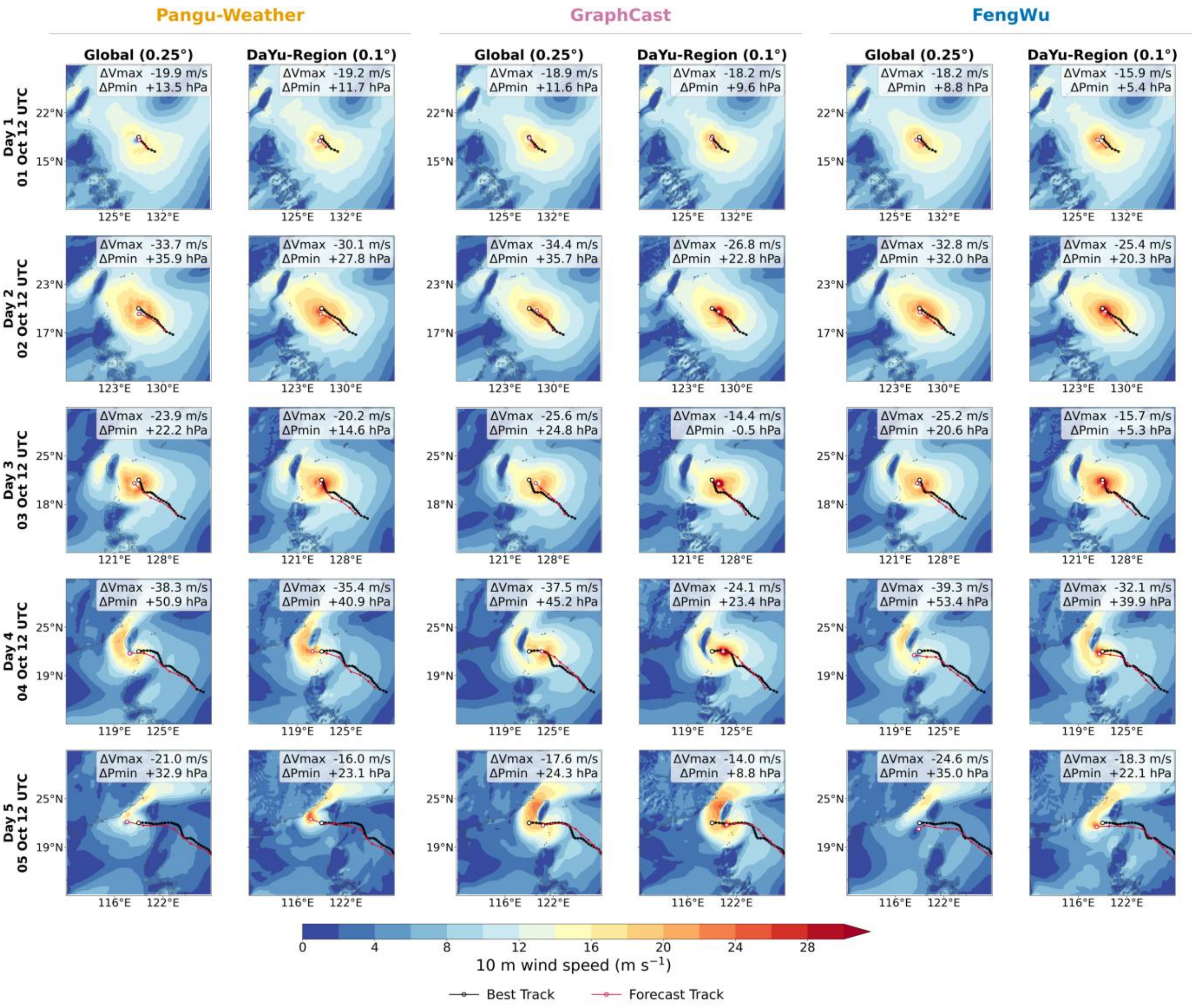

Supplementary Figure S11. Regional refinement of tropical cyclone Koinu under different global forecast drivers. Forecasts initialized at 12:00 UTC on 30 September 2023 are shown at lead times of 1–5 d (rows). Column pairs correspond to Pangu-Weather, GraphCast and FengWu, from left to right. Within each pair, the global forecast (0.25°) is shown on the left and the corresponding DaYu-Region forecast (0.1°) on the right. All regional forecasts use the same fixed weights trained with DaYu-Global backgrounds, without driver-specific fine-tuning. Shading indicates 10 m wind speed (m $s^{-1}$). Black and red lines denote the NOAA IBTrACS best track and forecast track, respectively; open circles mark the centres at the corresponding valid time. Annotations give ΔVmax and ΔPmin, defined as forecast minus best-track estimates of maximum 10 m wind speed (m $s^{-1}$) and minimum mean sea-level pressure (hPa), respectively. Negative ΔVmax and positive ΔPmin indicate underestimated intensity.

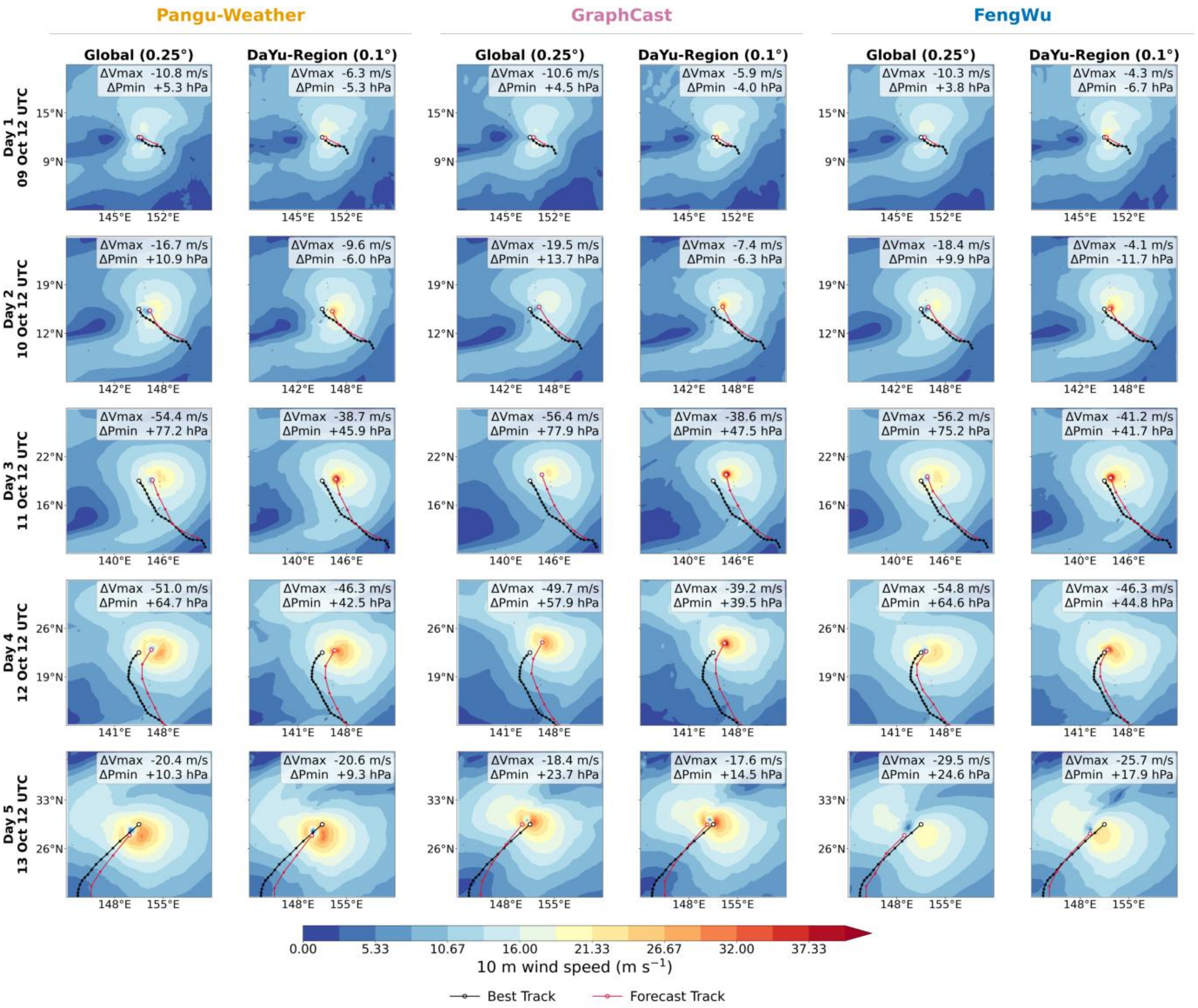

Supplementary Figure S12. Regional refinement of tropical cyclone Bolaven under different global forecast drivers. Forecasts initialized at 12:00 UTC on 8 October 2023 are shown at lead times of 1–5 d (rows). Column pairs correspond to Pangu-Weather, GraphCast and FengWu, from left to right. Within each pair, the global forecast (0.25°) is shown on the left and the corresponding DaYu-Region forecast (0.1°) on the right. All regional forecasts use the same fixed weights trained with DaYu-Global backgrounds, without driver-specific fine-tuning. Shading indicates 10 m wind speed (m $s^{-1}$). Black and red lines denote the NOAA IBTrACS best track and forecast track, respectively; open circles mark the centres at the corresponding valid time. Annotations give ΔVmax and ΔPmin, defined as forecast minus best-track estimates of maximum 10 m wind speed (m $s^{-1}$) and minimum mean sea-level pressure (hPa), respectively. Negative ΔVmax and positive ΔPmin indicate underestimated intensity.

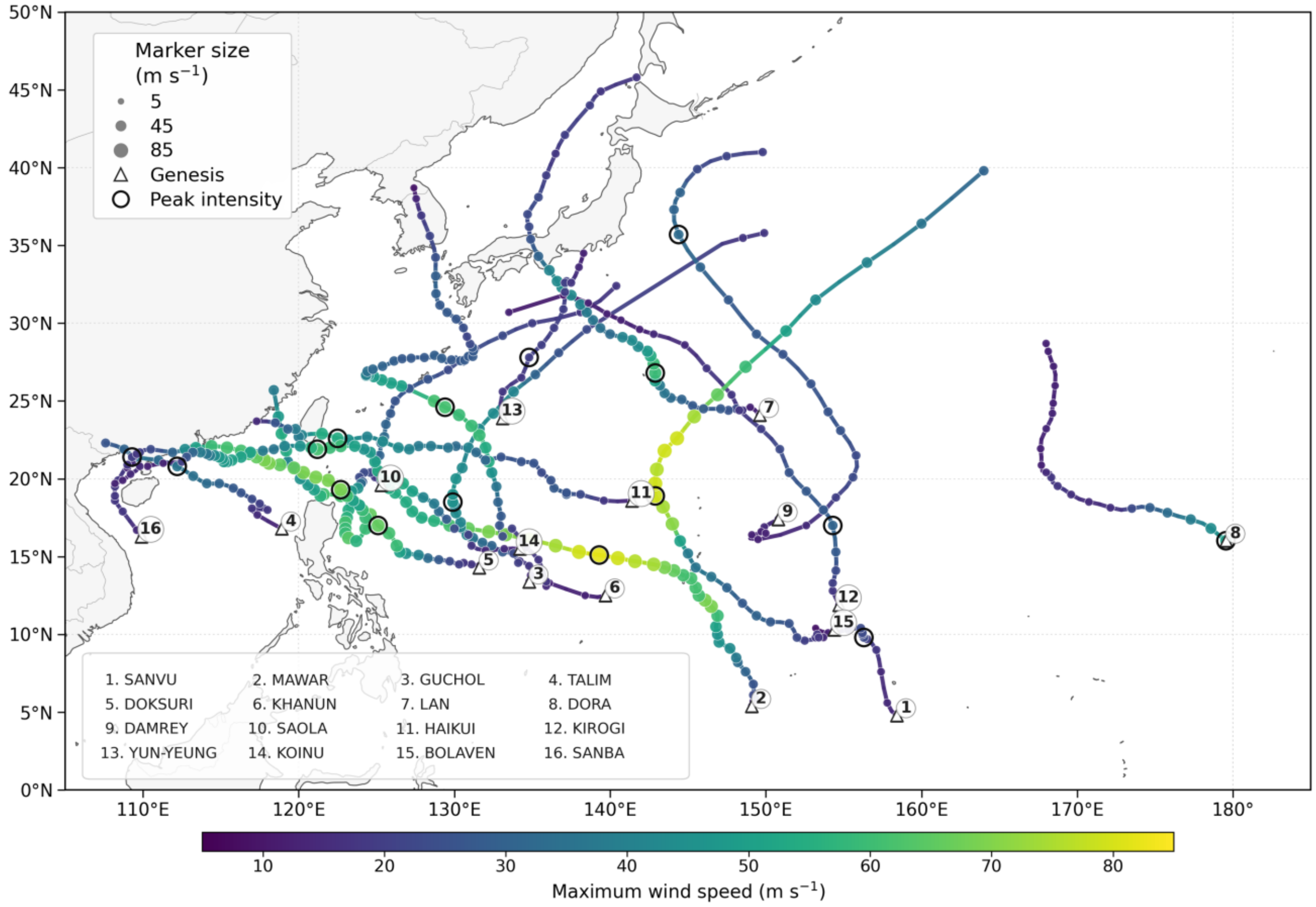

Supplementary Figure S13. Tracks and intensity evolution of the evaluated tropical cyclones in the western North Pacific in 2023. Tracks are shown for the 16 tropical cyclones included in the evaluation, using NOAA IBTrACS best-track records. Colours indicate maximum sustained wind speed (m $s^{-1}$) and marker sizes scale with wind speed. Triangles mark the first positions in the displayed records and open circles indicate the locations of peak intensity within those records. Numbers identify the individual tropical cyclones listed in the inset. The figure documents the geographic coverage and range of intensities represented in the tropical cyclone evaluation.

# Supplementary Methods

## Computational benchmarks

Supplementary Table S1. Computational resources, runtime and peak allocated GPU memory. DaYu results are measured for 120 h forecasts; the IFS-HRES literature estimate refers to 240 h [6].

| Model | Resource | Quantity | Forecast horizon (h) | Runtime | Peak allocated GPU memory (GiB) |
|---|---|---|---|---|---|
| DaYu-Global | NVIDIA A100 GPU | 1 | 120 | 23.6 s | 10.1 |
| DaYu-Region | NVIDIA A100 GPU | 1 | 120 | 27.9 s | 11.9 |
| DaYu-TC | NVIDIA A100 GPU | 1 | 120 | 47.3 s | 16.0 |
| IFS-HRES | CPU-based HPC cluster | 11,664 cores | 240 | ~60 min | — |

Inference is benchmarked over 20 successive 6 h steps (120 h) on a single NVIDIA A100 GPU. The complete DaYu-TC benchmark includes global inference, cropping and interpolation of global forecasts to the regional grid, and regional autoregressive prediction. Separate measurements quantify DaYu-Global and DaYu-Region. Runtime and peak allocated GPU memory are reported in Supplementary Table S1; the literature estimate for IFS-HRES concerns a 240 h global forecast and is provided as context.

## DaYu-TC framework

DaYu-TC is a global–regional nested data-driven forecasting framework (Fig. 1 in the main article). Its 0.25° global component, DaYu-Global, provides large-scale background information that evolves with forecast lead time. The global forecasts are cropped to the target region and interpolated to the regional grid to drive the 0.1° model, DaYu-Region. Incorporating global forecasts valid at future times provides constraints on large-scale circulation evolution beyond the information available from regional historical states alone.

## Dataset and preprocessing

### ERA5 data

DaYu-Global uses hourly ERA5 reanalysis on a 0.25° grid [30]. Each sample contains two input states separated by 6 h. The prediction target is the atmospheric state 6 h after the second input. A 1 h sliding sampling interval yields 280 500 training samples. The variables are listed in Supplementary Table S2. Data from 1990–2021 are used for training, 2022 for validation and 2023 for testing.

### High-resolution analysis data

DaYu-Region is trained and tested using operational analyses from the ECMWF Integrated Forecasting System (IFS) on a 0.1° grid [31]. The study domain covers the western North Pacific (WNP) (75.0°–204.6°E, 10.0°S–54.8°N). Data from 2016–2022 are used for training, with 2023 reserved for independent testing. Each regional sample contains two atmospheric states separated by 6 h, with the regional analysis 6 h after the later input as the prediction target. The regional model uses the same atmospheric variables as DaYu-Global, with a one-to-one correspondence between regional and global channels. Static predictors include terrain height and geographic position encodings derived from grid latitude and longitude, representing spatially varying surface forcing and domain geometry.

### Tropical cyclone best track data

Tropical cyclone (TC) positions and intensities are evaluated using the International Best Track Archive for Climate Stewardship (IBTrACS), provided by the US National Oceanic and Atmospheric Administration (NOAA) National Centers for Environmental Information (NCEI) [32,33]. Centre positions and valid times are read from the LAT, LON and ISO_TIME fields. Maximum sustained wind speed and minimum mean sea-level pressure are taken from USA_WIND and USA_PRES, respectively. USA_WIND contains US-agency 1-minute sustained wind estimates. The evaluation includes 16 TCs in the WNP in 2023; their tracks and intensity evolution are shown in Supplementary Fig. S13.

## DaYu-Global model structure

DaYu-Global is the global data-driven weather model developed in this study (Supplementary Fig. S14). The model takes two atmospheric states separated by 6 h as inputs, each with 69 channels. It predicts the change over the next 6 h, which is added to the current state to obtain the next forecast. Predicted states are then fed back as inputs for autoregressive forecasting.

DaYu-Global uses a U-shaped Transformer backbone. Surface and upper-air variables are encoded separately by patch embeddings and then fused. The encoder extracts local features through convolutional downsampling and residual blocks. The backbone consists of four Swin Transformer stages at the same feature resolution, using window attention and shifted-window interactions to establish spatial connections. The decoder restores spatial resolution and reconstructs forecast variables through upsampling and skip connections. Four enhancement modules strengthen the representation of weather evolution and extreme structures. The Trend Adapter developed here encodes differences between consecutive atmospheric states in separate branches and fuses the resulting features. These features are injected into the encoder through learnable gates to represent temporal changes in weather systems. Axial Attention establishes long-range connections along latitude and longitude, complementing window attention in representing large-scale spatial relationships. The Detail Adapter developed here recovers local information lost during downsampling. It extracts differences between high-resolution features and resampled low-resolution features. The resulting detail features are injected into the decoder through depthwise convolutions and gated residual connections to enhance wind-gradient and vortex representations. The extreme-weather Auxiliary Head provides additional supervision for shared features by predicting strong winds, low pressure and local pressure minima. This branch is used only during training.

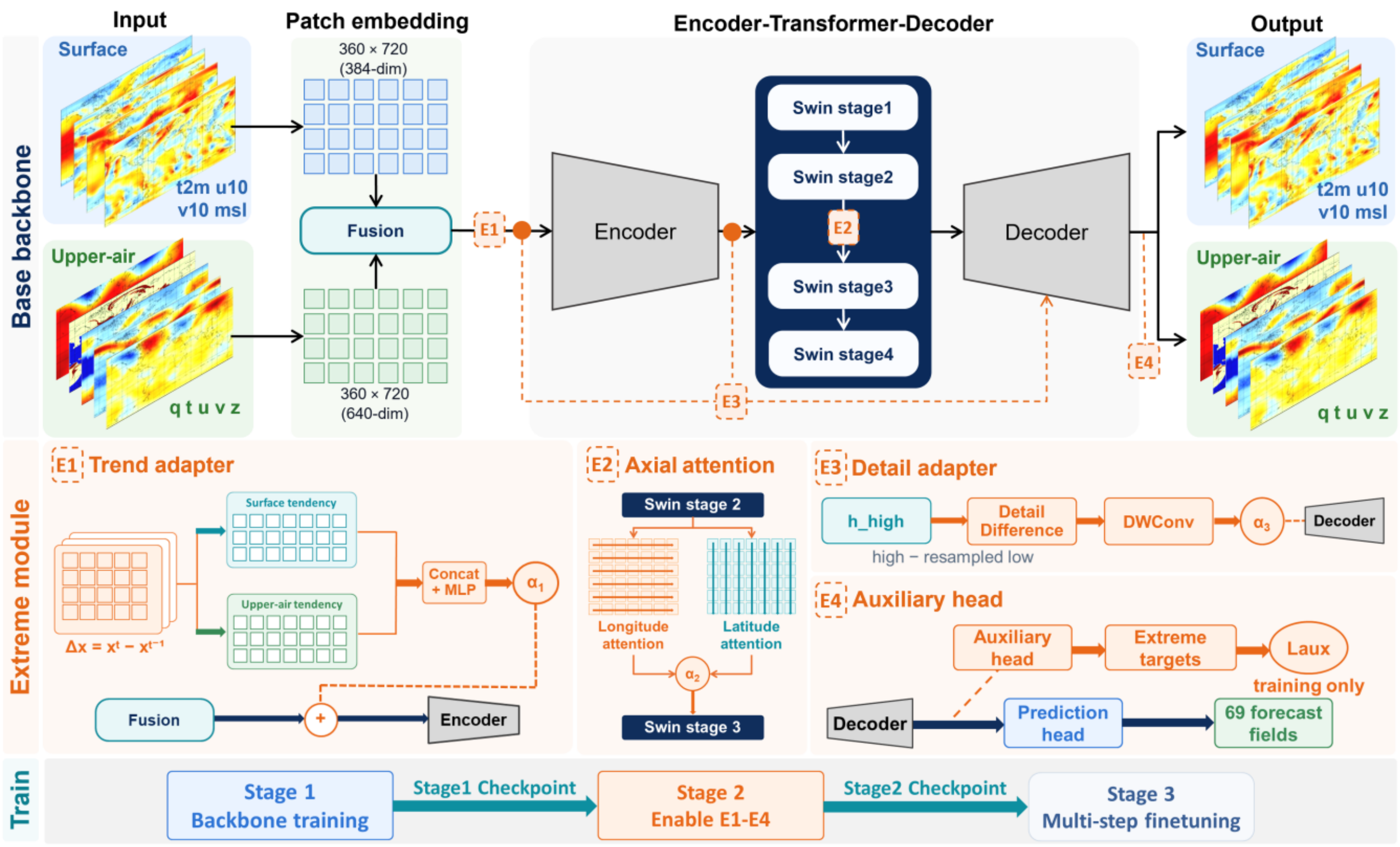

Supplementary Figure S14. DaYu-Global architecture and training workflow. Surface and upper-air variables are encoded separately, fused and processed by an encoder–Transformer–decoder backbone to predict the global atmospheric state 6 h ahead. The four enhancement modules are E1, Trend Adapter; E2, Axial Attention; E3, Detail Adapter; and E4, Auxiliary Head, which is used only during training. Training proceeds in three stages: backbone training, extreme-focused training with E1–E4 enabled and multi-step autoregressive fine-tuning. Each subsequent stage is initialized from the preceding stage's checkpoint. DWConv denotes depthwise convolution and MLP denotes a multilayer perceptron.

## DaYu-Region model structure

DaYu-Region provides 0.1° high-resolution forecasts over the WNP. It predicts the regional atmospheric state 6 h ahead from two consecutive regional states, a global forecast background valid at the target time and static geographic information. Regional historical fields provide information on local weather-system evolution. The global background is cropped and spatially interpolated to the 0.1° grid, supplying large-scale circulation constraints for the target time.

DaYu-Region uses the same surface–upper-air representations, U-shaped Transformer backbone and enhancement modules as DaYu-Global. It is initialized with the trained global weights. The two models differ primarily in spatial coverage, grid resolution and conditioning inputs. During multi-step forecasting, DaYu-Region advances autoregressively using two consecutive regional states and the global forecast valid at each target time. Together, the two models form the one-way nested DaYu-TC system.

## Training and inference

Training proceeds from global forecast learning to regional adaptation. DaYu-Global is trained first and then used to generate the global forecast backgrounds required for regional training. DaYu-Region is subsequently trained against 0.1° regional analyses. Actual global forecasts are used as background inputs during training, exposing DaYu-Region to the forecast errors present in its global driver.

### Loss function

DaYu-Global and DaYu-Region use the same types of training objectives. During basic training, the loss is the latitude-weighted mean absolute error (MAE) over all 69 forecast channels:

$$\mathcal{L}_G^{(1)} = \frac{1}{N}\sum_{v,i,j} w_i \left|\hat{y}_{v,i,j} - y_{v,i,j}\right|, \tag{S1}$$

where $\hat{y}$ and $y$ denote the predicted and reference fields, respectively; $v$ indexes atmospheric variables; $i, j$ are spatial grid indices; and $N$ is the total number of variable–gridpoint entries. Latitude weights are proportional to the cosine of latitude and normalized to have a mean of one:

$$w_i = \frac{\cos\phi_i}{\overline{\cos\phi_i}}, \quad \phi_i \in (-90°, 90°), \tag{S2}$$

where $\phi_i$ is the latitude of grid point $i$, and $\overline{\cos\phi_i}$ denotes the mean of $\cos\phi_i$ over all latitude grid points.

During extreme-focused training, a composite objective combines an extreme-aware field loss, a spatial-gradient loss, an extreme-value loss and an auxiliary-task loss:

$$\mathcal{L}_G^{(2)} = \mathcal{L}_{\text{field}}^{\text{ext}} + 0.03\mathcal{L}_{\text{grad}} + 0.02\mathcal{L}_{\text{ext}} + 0.02\mathcal{L}_{\text{aux}}. \tag{S3}$$

Here, $\mathcal{L}_{\text{field}}^{\text{ext}}$ is the extreme-aware field loss. It augments the basic latitude-weighted MAE with a continuous extreme score $S$, constructed jointly from strong winds, low pressure and local pressure minima in the reference fields. This assigns larger spatial loss weights to extreme-weather regions while retaining near-baseline weights elsewhere, increasing emphasis on structures associated with TCs and other high-impact weather. The spatial weights are expressed as:

$$W_{\text{ext}} = \frac{1+\alpha S}{\langle 1+\alpha S\rangle}. \tag{S4}$$

$\mathcal{L}_{\text{grad}}$ is applied only to the 10 m zonal and meridional wind components and mean sea-level pressure to strengthen the representation of sharp gradients:

$$\mathcal{L}_{\text{grad}} = \|\nabla_h\hat{y} - \nabla_h y\|_1 + \|\nabla_w\hat{y} - \nabla_w y\|_1. \tag{S5}$$

$\mathcal{L}_{\text{ext}}$ directly matches strong-wind and low-pressure extremes in physical units. It uses top-$K$ statistics to reduce instability caused by single-gridpoint displacements, with $K = 64$ in this study. $\mathcal{L}_{\text{aux}}$ is the loss for the E4 extreme-weather Auxiliary Head, which uses shared features to predict strong winds, low pressure and centre structures, providing supervision only during training.

These losses strengthen the representation of local gradients, anomalous states and extreme-weather structures while maintaining overall atmospheric-field forecast accuracy.

**Training settings**

Both models were trained using the AdamW optimizer [34] with $\beta_1 = 0.9$, $\beta_2 = 0.95$ and weight decay = 0.05. Training was performed on Ascend 910B neural processing units (NPUs), each with 64 GB of device memory. DaYu-Global and DaYu-Region used 32 and four NPUs, respectively. A batch size of one per device gave global batch sizes of 32 and 4.

In the first stage, the DaYu-Global backbone was trained for 25 epochs using the latitude-weighted MAE loss, with the enhancement modules frozen. The learning rate increased linearly from 0 to $5 \times 10^{-4}$ over 5000 warm-up steps, then decreased through cosine annealing to $1 \times 10^{-6}$. For the second stage, the model was initialized with the first-stage weights. The enhancement modules were activated and training continued for five epochs using the extreme-aware composite loss. Learning rates were $2 \times 10^{-5}$ for the backbone and $1 \times 10^{-4}$ for the enhancement modules, with 3000 warm-up steps followed by cosine annealing to a minimum learning rate of $1 \times 10^{-6}$. DaYu-Global then underwent five epochs of multi-step autoregressive fine-tuning. Following the replay-buffer strategy used in FengWu [35], model-predicted states were stored in a buffer and samples from different autoregressive steps were drawn with equal probability. The base learning rate was $2 \times 10^{-6}$ and decayed to $2 \times 10^{-8}$ through warm-up and cosine annealing.

DaYu-Region was trained in two stages, regional adaptation and extreme enhancement, each lasting five epochs. Initialized with DaYu-Global weights, the regional model first used the latitude-weighted MAE. Training then continued for five epochs using the extreme-aware composite loss. TC samples were resampled during this second stage to increase their representation in training. In each stage, the learning rate increased linearly to $1 \times 10^{-5}$ during warm-up and then decreased to 0 through cosine annealing. Training used 0.1° regional analyses as targets and DaYu-Global forecasts as background inputs, with a single-step 6 h prediction objective and no multi-step replay fine-tuning.

## Evaluation methods

Conventional atmospheric forecasts are evaluated using latitude-weighted RMSE (wRMSE) and latitude-weighted anomaly correlation coefficient (wACC), following WeatherBench 2. These metrics quantify forecast error and spatial agreement between anomalies, respectively [36]. The wRMSE is defined as:

$$\text{wRMSE} = \sqrt{\frac{\sum_{i,j} w_i \, (\hat{y}(x) - y(x))^2}{\sum_{i,j} w_i}}, \tag{S6}$$

where $\hat{y}$ and $y$ denote predicted and reference values, respectively, and the latitude weights $w_i$ follow Eq. (S2).

To assess large-scale anomaly patterns and their spatial distribution, wACC compares forecast and reference anomalies relative to climatology. It is defined as:

$$\text{wACC} = \frac{\sum_{i,j} w_i(\hat{y}_{v,i,j} - c_{v,i,j})(y_{v,i,j} - c_{v,i,j})}{\sqrt{\sum_{i,j} w_i(\hat{y}_{v,i,j} - c_{v,i,j})^2} \sqrt{\sum_{i,j} w_i(y_{v,i,j} - c_{v,i,j})^2}}, \tag{S7}$$

where $c_{v,i,j}$ is the climatological mean for the variable at the corresponding valid time. We use a daily climatology constructed by day of year (DOY) from 1993–2016 data. Values closer to 1 indicate greater spatial agreement between forecast and reference anomaly fields.

TC track and intensity skill is assessed using rolling forecasts for 16 TCs in the WNP in 2023. Centres are identified from forecast sea-level pressure and 10 m winds. The forecast centre is identified as the minimum-pressure location within 200 km of the previous centre. Maximum 10 m wind speed is calculated within 400 km of the diagnosed centre as an intensity metric.

Track error is measured by the Euclidean distance:

$$D_{\text{euc}} = \sqrt{(\Delta x)^2 + (\Delta y)^2}, \tag{S8}$$

where $\Delta x$ and $\Delta y$ are the meridional and zonal distance differences, respectively.

TC intensity is characterized by maximum 10 m wind speed and minimum mean sea-level pressure, with RMSE calculated against best-track estimates for each metric. Maximum 10 m wind speed is obtained from the 10 m zonal and meridional wind components:

$$\text{MWS} = \max_x \sqrt{U_{10}(x)^2 + V_{10}(x)^2}, \tag{S9}$$

where $U_{10}(x)$ and $V_{10}(x)$ are the 10 m zonal and meridional wind components, respectively, at location $x$.

### Overall intensity RMSE and cross-driver evaluation

Lead-specific RMSE is calculated from valid forecast–best-track pairs at each lead time. Overall RMSE is calculated as the square root of the sum of squared errors across all included lead times divided by the total number of valid samples. Equivalently, it is the square root of the sample-count-weighted mean of the squared lead-specific RMSEs. This differs from an unweighted average of RMSE across lead times. Percentage reductions are calculated as 100 × (RMSEbaseline − RMSEnested)/RMSEbaseline, using either the relevant global driver or IFS-HRES as the baseline specified in each comparison.

For cross-driver experiments, the same DaYu-Region weights trained with DaYu-Global backgrounds are retained. Forecasts from Pangu-Weather, GraphCast and FengWu are cropped and interpolated to the regional grid to replace the DaYu-Global backgrounds at the corresponding target times. The regional model then advances its own states autoregressively without retraining or fine-tuning. Maximum-wind and minimum-pressure errors are evaluated separately against the same IBTrACS best-track estimates.

### Global extreme-anomaly evaluation

To assess the prediction and preservation of extreme weather anomalies, we define extreme events using local calendar-day percentile thresholds. A local 95th-percentile (P95) threshold is calculated for each calendar day using ERA5 data from 1993–2016. Following the centred 5-day-window approach used for percentile-based extreme indices [37], samples at 00:00, 06:00, 12:00 and 18:00 UTC are pooled from the target calendar day and the two days on either side. Each gridpoint threshold is therefore estimated from 24 × 5 × 4 = 480 historical samples.

For the 10 m and 850 hPa zonal and meridional wind components, the extreme-anomaly magnitude is defined as:

$$\mathrm{A} = |X - \mathrm{C}|, \tag{S10}$$

where $X$ is the atmospheric field and $\mathrm{C}$ is the ERA5 climatology for the corresponding calendar day. For mean sea-level pressure, only anomalously low pressure is considered, with the anomaly defined as:

$$\mathrm{A} = \mathrm{C} - \mathrm{MSLP}. \tag{S11}$$

The P95 of historical anomaly magnitudes is then calculated at each grid point, yielding a threshold Q95(d, i, j) that varies spatially and seasonally. During verification, a grid point is classified as an extreme event when the ERA5 anomaly reaches or exceeds its local P95 threshold.

To quantify anomaly-amplitude preservation at the locations of reference extreme events, we calculate the relative mean bias error (rMBE) over ERA5-defined extreme grid points. The relative mean bias is calculated from extreme-anomaly amplitudes:

$$\mathrm{rMBE} = 100 \times \frac{\sum_i w_i M_i (A_{f,i} - A_{o,i})}{\sum_i w_i M_i A_{o,i}} \ (\%), \tag{S12}$$

where $A_f$ and $A_o$ denote forecast and ERA5 anomaly amplitudes, respectively; $M_i$ is the extreme-event mask defined by the ERA5 P95 thresholds; and $w_i$ is the latitude weight. The ideal rMBE is 0. Negative values indicate systematic attenuation of extreme-anomaly amplitudes at ERA5 extreme locations, whereas positive values indicate systematic amplification.

Supplementary Table S2. Forecast variables and channel configuration of DaYu-Global and DaYu-Region.

| **Category** | **Variable** | **Symbol** | **Vertical level(s)** |
|---|---|---|---|
| Near-surface | 2 m temperature | T2M | 2 m |
| | 10 m zonal wind | U10 | 10 m |
| | 10 m meridional wind | V10 | 10 m |
| | Mean sea-level pressure | MSLP | Sea level |
| Upper-air | Specific humidity | Q | 13 pressure levels |
| | Temperature | T | 13 pressure levels |
| | Zonal wind | U | 13 pressure levels |
| | Meridional wind | V | 13 pressure levels |
| | Geopotential | Z | 13 pressure levels |

The 13 pressure levels are ordered as 50, 100, 150, 200, 250, 300, 400, 500, 600, 700, 850, 925 and 1000 hPa.

## References cited in the Supplementary Data

[6] Lam R, Sanchez-Gonzalez A, Willson M, et al. Learning skillful medium-range global weather forecasting. Science 2023; 382: 1416–21. doi: 10.1126/science.adi2336.

[30] Hersbach H, Bell B, Berrisford P, et al. The ERA5 global reanalysis. Q J R Meteorol Soc 2020; 146: 1999–2049. doi: 10.1002/qj.3803.

[31] ECMWF. IFS Documentation CY47R3—Part II: Data assimilation. Reading: ECMWF, 2021. doi: 10.21957/t445u8kna.

[32] Knapp KR, Kruk MC, Levinson DH, et al. The International Best Track Archive for Climate Stewardship (IBTrACS): Unifying tropical cyclone best track data. Bull Am Meteorol Soc 2010; 91: 363–76. doi: 10.1175/2009BAMS2755.1.

[33] Gahtan J, Knapp KR, Schreck CJ, et al. International Best Track Archive for Climate Stewardship (IBTrACS) Project, Version 4r01. NOAA National Centers for Environmental Information, 2024. doi: 10.25921/82ty-9e16.

[34] Loshchilov I, Hutter F. Decoupled weight decay regularization. In: Proceedings of the International Conference on Learning Representations, New Orleans, LA, 6–9 May 2019. https://openreview.net/forum?id=Bkg6RiCqY7

[35] Chen K, Han T, Ling F, et al. The operational medium-range deterministic weather forecasting can be extended beyond a 10-day lead time. Commun Earth Environ 2025; 6: 518. doi: 10.1038/s43247-025-02502-y.

[36] Rasp S, Hoyer S, Merose A, et al. WeatherBench 2: A benchmark for the next generation of data-driven global weather models. J Adv Model Earth Syst 2024; 16: e2023MS004019. doi: 10.1029/2023MS004019.

[37] Zhang X, Hegerl G, Zwiers FW, et al. Avoiding inhomogeneity in percentile-based indices of temperature extremes. J Clim 2005; 18: 1641–51. doi: 10.1175/JCLI3366.1.